\documentclass[aps,
prl,
twocolumn,
reprint,
noeprint,
superscriptaddress,
amsmath,
amssymb,
]{revtex4-2}
\usepackage[english]{babel}
\usepackage{amsmath}

\usepackage{graphicx}
\usepackage{color}
\usepackage{bm,bbold}
\usepackage{braket}

\usepackage[colorlinks=true, allcolors=blue]{hyperref}
\usepackage{times}
\usepackage[dvipsnames]{xcolor}

\usepackage{cleveref}

\newcommand\LHVMeq{\mathrel{\stackrel{\makebox[0pt]{\mbox{\normalfont\tiny LHVM}}}{=}}}

\newcommand{\ie}{\textit{i.e.} }

\newcommand{\rev}[1]{{\color{black}#1}}

\begin{document}

\title{Observing Bell Inequality Violation Beyond the Qubit Bound in a Spinor Bose--Einstein Condensate}

\author{Wenxin Xu}
\thanks{These authors contributed equally to this work.}
\affiliation{State Key Laboratory of Low Dimensional Quantum Physics, Department of Physics, Tsinghua University, Beijing 100084, China}

\author{Junshuang Hu}
\thanks{These authors contributed equally to this work.}
\affiliation{State Key Laboratory of Low Dimensional Quantum Physics, Department of Physics, Tsinghua University, Beijing 100084, China}

\author{Guillem M\"uller-Rigat}
\affiliation{Faculty of Physics, Astronomy and Applied Computer Science, Jagiellonian University, ul. \L ojasiewicza 11, 30-348 Kraków, Poland.}

\author{Xinwei Li}
\affiliation{Beijing Academy of Quantum Information Sciences, Beijing 100193, China}

\author{Qi Liu}
\email[]{qiliu23@mit.edu}
\affiliation{National Laboratory of Solid State Microstructures and School of Physics,
Collaborative Innovation Center of Advanced Microstructures, Nanjing University, Nanjing 210093, China}

\author{Matteo Fadel}
\email[]{fadelm@phys.ethz.ch}
\affiliation{Department of Physics, ETH Zürich, 8093 Zürich, Switzerland}

\author{Li You}
\email[]{lyou@mail.tsinghua.edu.cn}
\affiliation{State Key Laboratory of Low Dimensional Quantum Physics, Department of Physics, Tsinghua University, Beijing 100084, China}
\affiliation{Beijing Academy of Quantum Information Sciences, Beijing 100193, China}
\affiliation{Frontier Science Center for Quantum Information, Beijing, China}
\affiliation{Hefei National Laboratory, Hefei, Anhui 230088, China}

\date{\today}
\begin{abstract}
Correlations allowed by quantum mechanics can defy any classical explanation, with Bell nonlocality standing as their most profound and operationally powerful manifestation. While nonlocality has been demonstrated across a wide range of platforms, in many-body systems it has remained limited to ensembles of qubits, leaving the observation of higher-dimensional multipartite Bell correlations an open challenge. 
Here we report the observation of qutrit Bell correlations in a spin-1 $^{87}$Rb Bose-Einstein condensate via the violation of a Bell witness based on collective spin observables only. Exploiting spin-exchange collisions in an ensemble of $N \simeq 3.1 \times 10^{4}$ atoms, we generate spin-nematic squeezing of $-11.8(7)$ dB and observe a violation that surpasses the minimum bound achievable by a collection of $N$ qubits, providing direct evidence of genuine multipartite qutrit Bell correlations. Our results establish spinor Bose-Einstein condensates as a viable platform for investigating high-dimensional Bell correlations in the many-body regime, and demonstrate that coarse-grained collective measurements suffice to certify the dimensionality of quantum correlations at macroscopic scales.

\end{abstract}
\maketitle

\textit{Introduction}---Bell nonlocality~\cite{brunner_bell_2014} plays a foundational role in quantum physics, representing correlations that admit no classical explanation within any local hidden-variable framework. Since Bell's seminal work~\cite{bell_einstein_1964}, the violation of Bell inequalities has served as the definitive signature of quantum correlations even stronger than entanglement, with far-reaching implications for both the foundations of quantum mechanics and its technological applications, including device-independent quantum key distribution~\cite{ekert_quantum_1991, barrett_no_2005, acin_device-independent_2007} and certified randomness generation~\cite{pironio_random_2010, colbeck_private_2011}. Crucially, nonlocality constitutes not merely a conceptual curiosity but also an operationally meaningful resource that can be certified directly from measurement statistics, independently of assumptions about the underlying physical realization. 

Over the past decades, remarkable progress has been made in demonstrating Bell nonlocality across a broad range of physical platforms. Early experiments focused on bipartite systems~\cite{hofmann_heralded_2012,pfaff_demonstration_2013,ansmann_violation_2009}, while subsequent efforts extended these ideas to multipartite settings~\cite{lanyon_experimental_2014,eibl_experimental_2003,zhao_experimental_2003}, where nonlocal correlations acquire a much richer and more complex structure. In particular, multipartite qubit systems have enabled the observation of Bell inequality violations associated with highly entangled states, such as Greenberger–Horne–Zeilinger states, realized in platforms ranging from photonic networks~\cite{zhong_12-photon_2018} to superconducting quantum processors~\cite{song_generation_2019}. Parallel developments in atomic physics have shown that Bell correlations can persist even in the many-body regime~\cite{tura_detecting_2014}: by exploiting collective spin observables, Bell correlations have been observed in atomic ensembles containing up to several hundred thousand particles~\cite{schmied_bell_2016,Kasevich2017PRL}, without the need for single-particle control or measurement resolution. 

\begin{figure}[!t]
    \centering
    \includegraphics[width=1\columnwidth]{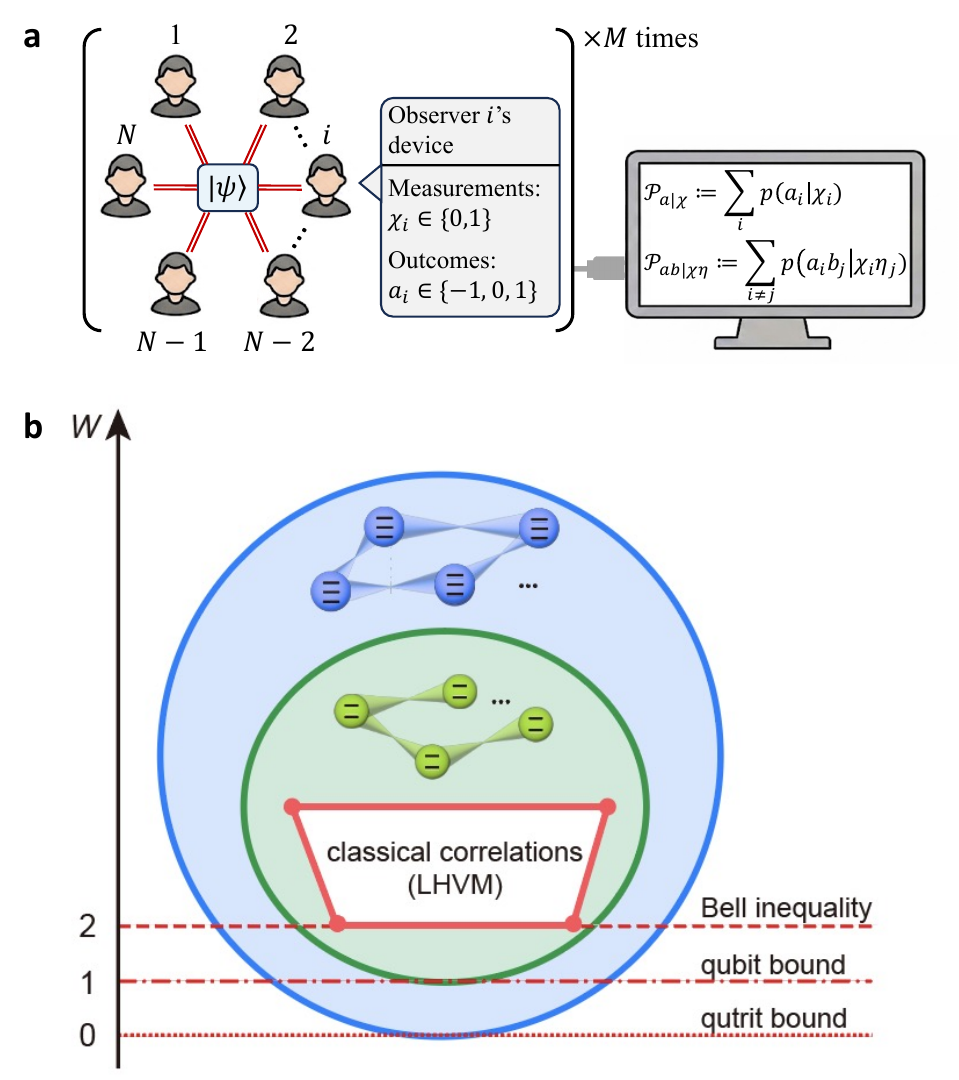}
    \caption{\textbf{High-dimensional Bell correlation detection.} \textbf{a}. $N$ observers perform binary local measurements $\chi_i \in \{0,1\}$ on a shared physical system with ternary outcomes $a_i \in \{-1,0,1\}$. The measurements are repeated multiple times to yield one- and two-body conditional probabilities $\mathcal{P}_{a|\chi}$ and $\mathcal{P}_{ab|\chi\eta}$, which are used in the Bell inequalities. \textbf{b}. Dimension-based hierarchy of quantum correlations. Classical correlations, constrained by local hidden variable models (LHVMs), define a polytope bounded by local deterministic strategies, whose facets formulate Bell inequalities. Quantum states of different Hilbert space dimensionality can allow for only down to some level of violation. Therefore, it is possible to construct criteria able to certify the system's dimensionality according to the experimentally observed Bell violation. The green and blue shading indicate the regimes accessible with qubit and qutrit correlations, associated with the corresponding bounds of Bell witness $W=1$ and $W=0$, respectively.}
    \label{fig:1}
\end{figure}

Despite these advances, existing many-body demonstrations of Bell correlations have been limited to two-level (qubit) descriptions. While qubit systems already exhibit rich correlation structures, their intrinsically two-dimensional local Hilbert spaces constrain the dimensionality accessible to device-independent certification. This restriction is not merely technical: higher-dimensional quantum systems (qudits) support forms of correlations inaccessible to qubits, yield stronger violations of certain Bell inequalities~\cite{GuillemPRXQuantum,kobus_multisetting_2025}, and enable the certification of genuinely high-dimensional entanglement~\cite{muller-rigat_three-outcome_2026}. More broadly, extending Bell correlations to multipartite qudit systems provides new opportunities for testing quantum mechanics, probing the structure of many-body correlations, and enhancing quantum information protocols beyond qubit-based limits.

Spinor Bose–Einstein condensates~\cite{Law, Chang2025NP} offer an especially compelling platform for pursuing this goal. In these systems, atoms possess multiple internal spin states that can be coherently coupled and entangled through well-controlled interactions. This combination of large particle number, collective addressability, and intrinsic multilevel structure makes spinor condensates ideally suited for exploring high-dimensional nonlocal correlations in the many-body regime. In particular, collision-based interaction dynamics naturally generate spin-nematic squeezing~\cite{Hamley2012}, producing strongly correlated states that allow for accessing genuinely higher-dimensional quantum resources and open up diverse applications, including entanglement-enhanced sensing of scalar~\cite{mao2023quantum} or vector~\cite{cao_joint_2025} electromagnetic field, as well as atomic gravimeter~\cite{PhysRevX.15.011029}.

In this work, we demonstrate Bell correlations beyond the qubit paradigm using a spin-1 Bose–Einstein condensate of rubidium-87 atoms. We show that spin-nematic squeezed states prepared in an ensemble of $N\simeq3.1\times10^{4}$ atoms can violate a Bell-correlation witness by an amount that cannot be achieved by any state of $N$ qubits. This violation indicates that the observed correlations must involve genuine three-level systems, thereby witnessing multipartite qutrit Bell correlations in a macroscopic quantum system. Our results establish spinor condensates as a powerful platform for the study of high-dimensional Bell correlations in the many-body regime, thus providing new tools for probing the limits of quantum theory in large, interacting systems. 

\textit{High-dimensional Bell correlations}---We consider a Bell experiment where $N$ observers, labeled by $i \in\{1, \ldots, N\}$, perform measurements on a physical system they share (see Fig.~\ref{fig:1}\textbf{a}). Each observer can choose to perform one out of two possible local measurements $\chi_i \in \{0,1\}$ and each measurement will yield one of three possible outcomes $a_i \in \{-1,0,1\}$.
The complete Bell experiment is carried out over many repetitions. In each run, a new copy of the physical system is distributed to the parties, who independently select a measurement to perform on their respective subsystems and record the corresponding outcomes. Crucially, the choices of measurements are statistically independent of the system state, and no communication between the parties is permitted during the experiment.

After a sufficiently large number of runs, one can estimate the conditional probability distribution $p(\bm{a}|\bm{\chi})$ associated with the observed data. 
This distribution specifies the probability of obtaining outcomes $\bm{a}:=(a_1,a_2,\ldots,a_N)$ given that the measurement settings $\bm{\chi}:=(\chi_1,\chi_2,\ldots,\chi_N)$ are chosen.
Classical correlations imply that $p(\bm{a}|\bm{\chi})$ can be explained by pre-established agreements, \ie by a local hidden variable model (LHVM).
This means that $p(\bm{a}|\bm{\chi}) \;\;\LHVMeq\; \int p(\lambda) p(a_1|\chi_1,\lambda)...p(a_N|\chi_N,\lambda) \mathrm{d}\lambda$,  where $p(\lambda)$ is the probability of using agreement $\lambda$ \cite{bell_einstein_1964}. If this is not the case, namely if the observed statistics $p(\bm{a}|\bm{\chi})$ cannot be written according to a LHVM, we have to conclude the presence of Bell nonlocality in the system.

Concretely, Bell correlation is revealed through the violation of a Bell inequality, that is, a constraint that must be satisfied by all LHVMs. In multipartite scenarios, however, identifying such inequalities becomes exceedingly challenging due to the rapidly increasing complexity of the correlations involved. Consequently, it is often necessary to introduce simplifications to the problem, for instance by imposing physically motivated symmetries that reduce the space of correlations to be explored~\cite{tura_detecting_2014}. 
Importantly, these symmetries do not enter as an additional assumption on the system, but just lead to inequalities especially tailored to those situations. 
In addition, another common simplification consists of truncating the space of correlations to low-order collective observables that are practically accessible in experiments~\cite{frerot_probing_2023}, although such Bell correlation measurements do not demonstrate loophole-free Bell nonlocality, owing to the absence of spatial separation and freedom of choice.

In the following, we consider permutationally invariant (PI) one- and two-body conditional probabilities defined by the observables $\mathcal{P}_{a|\chi} := \sum\limits_{i} p(a_i|\chi_i)$ and $\mathcal{P}_{ab|\chi\eta} := \sum\limits_{i\neq j} p(a_i b_j|\chi_i \eta_j)$.
We further restrict our analysis to Bell inequalities invariant under relabeling of inputs $\{0,1\}$ and outputs $\{-1,+1\}$. This three-outcome structure naturally matches our spin-1 system, where spin-exchange collisions coherently couple the Zeeman states through $\ket{0}\ket{0}\leftrightarrow\ket{-1}\ket{+1}$, and population measurements resolve the three outcomes $m_F = -1,0,+1$\cite{Hamley2012,KitzingerPRA21,Luo17deterministic}.

Having specified the set of observables and symmetries, Bell inequalities can now be systematically derived \cite{tura_detecting_2014,AnnPhys,GuoPRL23,aloy_deriving_2024}.
The strategy consists in using the fact that the set of LHVM correlations defines a polytope, whose vertices are determined by local deterministic strategies that can be listed (see Fig.~\ref{fig:1}\textbf{b}).
From these vertices one then uses algorithms to obtain hyperplanes defining the polytope facets, which consist of tight Bell inequalities.
Carrying out this procedure we obtain the PI Bell inequality
\begin{equation}\label{eq:5dimPIBI}
    B = \tilde{\mathcal{P}}_{\alpha}  +  \tilde{\mathcal{P}}_{\alpha\alpha} - 2 \tilde{\mathcal{P}}_{\alpha\beta} \geq 0 \;,
\end{equation}
where $\tilde{\mathcal{P}}_{\alpha} := \mathcal{P}_{-1|0}+\mathcal{P}_{-1|1} + \mathcal{P}_{1|0}+\mathcal{P}_{1|1}$, $\tilde{\mathcal{P}}_{\alpha\alpha}  := \mathcal{P}_{-1-1|00} + \mathcal{P}_{-1-1|11} + \mathcal{P}_{11|00} + \mathcal{P}_{11|11}$, and $\tilde{\mathcal{P}}_{\alpha\beta} :=  \mathcal{P}_{-11|01} + \mathcal{P}_{1-1|01}$ are one- and two-body observables invariant under relabeling $\pm 1$ (see the Supplemental Material (SM) \cite{SupplementalMaterial}).
If a system exhibits statistics such that $B < B_{\rm local}=0$, then it cannot be explained by a LHVM and Bell nonlocality is detected.
At this point, let us emphasize that inequality~\eqref{eq:5dimPIBI} can be violated even by qubit states. Remarkably, for a $N$-particle system with $n\leq N$ particles restricted to qubits, we derive the dimension-dependent bound~\cite{SupplementalMaterial}
\begin{equation}\label{qutrit}
B \geq -n/4 - (N-n)/2.
\end{equation}
Specifically, $n=N$ qubits can violate inequality~\eqref{eq:5dimPIBI} only down to $B_{\rm qubit}=-N/4$, whereas quantum mechanics predicts a lower bound $B_{\rm qutrit}=-N/2$ for the all-qutrit case ($n=0$). Thus, our Bell inequality can also be employed to certify the number of qutrits in the system. Indeed, the presence of at least $N-n+1$ qutrits can be certified if \eqref{qutrit} is violated. 

Spinor BECs further provide access to ensembles of thousands of particles~\cite{Law,Chang2025NP}, albeit without single-particle addressability and hence with operations and measurements restricted to collective degrees of freedom. In our case of interest, the latter is defined by associating to each particle a spin-1 angular momentum operator $\hat l_\mu^{(i)}$, with $\mu\in\{x,y,z\}$ and $i$ the particle label, and then introducing the collective spin operator $\hat L_\mu = \sum_{i=1}^N \hat l_\mu^{(i)}$.
Assuming that measurement outcomes probabilities can be described by measurements of these collective operators on a quantum state according to Born's rule, inequality~\eqref{eq:5dimPIBI} can be expressed as the expectation value of a Bell witness operator. 
Particularly convenient operators are obtained from choosing measurement settings to be identical on each subsystem, as this leads to witnesses based on collective operators.

We introduce the witness operator
\begin{equation}\label{W}
\hat{W}=\underbrace{\frac{3}{N}\left(\hat{Q}_{y z}^2+\hat{Q}_{x z}^2\right)}_{\hat{W}_{\rm non}}+\underbrace{\frac{5}{2 N} \hat{Q}_{z z}+\frac{10}{3}}_{\hat{W}_{\rm lin }},
\end{equation}
which contains the nonlinear term $\hat{W}_{\rm non}$ and the linear term $\hat{W}_{\rm lin}$. Here, we have used the collective nematic tensor operator $\hat Q_{\mu\nu}= \sum_{i=1}^N \left[ \hat l_\mu^{(i)} \hat l_\nu^{(i)}+\hat l_\nu^{(i)}\hat l_\mu^{(i)} - \frac{4}{3}\delta_{\mu\nu} \right]$~\cite{Hamley2012}, and prove that from the dimension bounds we get $W=\langle \hat W \rangle=4B/N+2$~\cite{SupplementalMaterial}. Thus, $W<2$ reveals Bell correlations, while $W<1$ further violates the qubit bound and thus certifies that these correlations must involve genuine qudits. In the following, we discuss the test of this witness in a spinor BEC experiment.

\begin{figure}[!ht]
    \centering
    \includegraphics[width=0.98\linewidth]{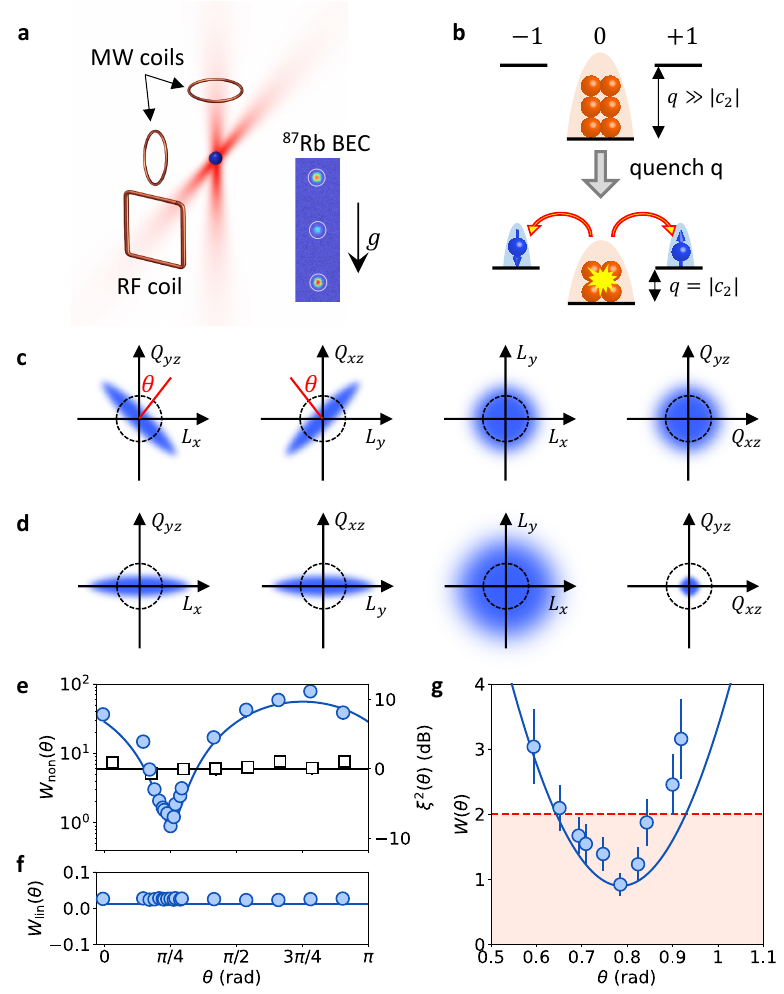}
    \caption{\textbf{Observation of Bell correlations in a spin-1 BEC.} \textbf{a}. Illustration of the experimental system. A spinor BEC is tightly confined in a crossed optical dipole trap, with its spin dynamics controlled by microwave and RF fields. During time of flight, the three spin components are spatially separated by a Stern--Gerlach gradient along the direction of gravity and subsequently detected by absorption imaging. \textbf{b}. Illustration of spin-mixing dynamics, which transfers atom pairs from $|0\rangle$ to $\ket{\pm1}$ and generates spin-nematic squeezing. \textbf{c}. Husimi distributions of the spin-nematic squeezed state, revealing anisotropic quantum fluctuation in the spin-nematic planes $L_x$-$Q_{yz}$ and $L_y$-$Q_{xz}$ and isotropic noise in the spin plane $L_x$-$L_y$ and nematic plane $Q_{xz}$-$Q_{yz}$. The dashed circle radius corresponds to the quantum projection noise (QPN) of the polar state. \textbf{d}. Accumulation of spinor phases rotates the state in the spin-nematic planes, while contracting or dilating the distributions in the other two planes. \textbf{e}. Measured nonlinear term of the Bell witness, $W_{\rm non}$, versus orientation angle $\theta$ for the spin-nematic squeezed state (blue circles) and the polar state (white squares). The right $y$-axis shows the corresponding squeezing parameter $\xi^2(\theta)$. \textbf{f}. Linear term of the Bell witness, $W_{\rm lin}$, for the spin-nematic squeezed state. \textbf{g}. Complete Bell witness, which maximally violates the Bell inequality (dashed line) at $\theta=0.79$ rad. The orange shading indicates the regime where Bell correlations are detected. All data points in this manuscript are averaged over at least $50$ shots with error bars standing for $1\sigma$ statistical uncertainty. Blue solid curves in \textbf{e-g} are numerical simulation results with the truncated Wigner method \cite{SupplementalMaterial}.}
    \label{fig:2}
\end{figure}

\textit{Spin-nematic squeezing in spin-1 BECs}---We consider a spin-1 $^{87}$Rb BEC in the $F=1$ hyperfine ground-state manifold with three Zeeman sublevels $\left|m_F\right\rangle$ ($m_F\in\{0,\pm1\}$) confined tightly in a crossed optical dipole trap (see Fig.~\ref{fig:2}\textbf{a}). Within the single-mode approximation~\cite{Law,Yi}, the internal spin dynamics is governed by the Hamiltonian

\begin{equation}\label{Ham}
\begin{split}
\hat{H} = \frac{c_2}{2N} \Big[
&2(\hat{a}_1^\dagger \hat{a}_{-1}^\dagger \hat{a}_0 \hat{a}_0 + \text{h.c.}) \\
&+ (2\hat{N}_0 - 1)(N - \hat{N}_0)
\Big] - q\hat{N}_0 .
\end{split}
\end{equation}
where $\hat{a}_{m_F}^\dagger$~($\hat{a}_{m_F}$) denotes the creation~(annihilation) operator for the $m_F$ spin component, $\hat{N}_{m_F}=\hat{a}_{m_F}^\dagger\hat{a}_{m_F}$ its atom number, $c_2$ the spin-mixing interaction strength, $N$ the total atom number, and $q$ the quadratic Zeeman shift (QZS). 
The term in the square bracket coincides with $\hat{\mathbf{L}}^2$, and is written as the sum of two contributions.
The first describes the spin-mixing dynamics, which creates (annihilates) pairs of $\ket{\pm1}$ atoms from (into) $\ket0$ (see Fig.~\ref{fig:2}\textbf{b}), while the second one denotes the energy shift due to the elastic atomic collisions. 
The linear term describing the QZS is tunable with a bias magnetic field or off-resonant microwave dressing~\cite{Gerbier,Jiang}. 

\begin{figure}[!t]
    \centering
    \includegraphics[width=1\linewidth]{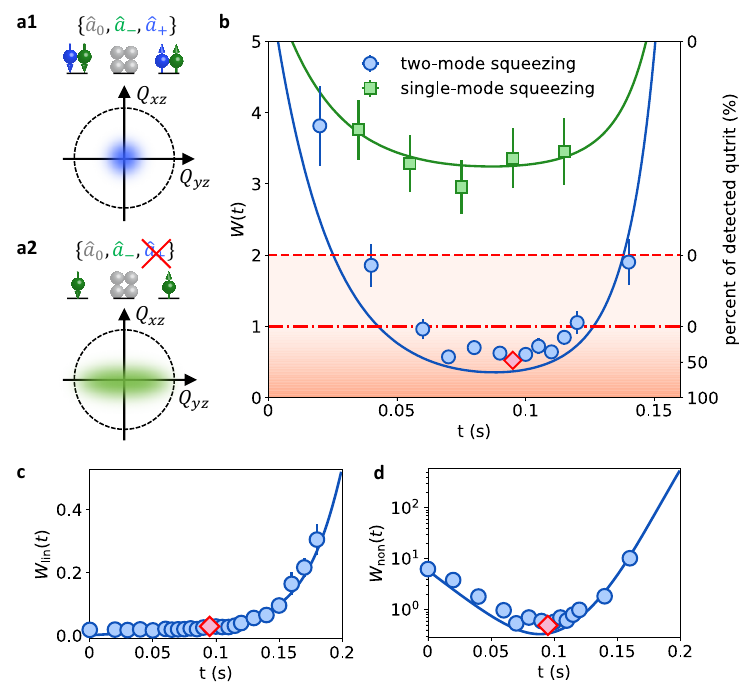}
    \caption{\textbf{Violation of Bell inequality beyond the qubit bound.} \textbf{a1}. Spin-nematic squeezed state with the optimal squeezing axis aligned along ${Q}_{xz/yz}$ exhibits an isotropic and contracted distribution in the ${Q}_{yz}$-${Q}_{xz}$ plane, revealing the simultaneous squeezing of two modes $\hat{a}_{\pm}=(\hat{a}_1\pm\hat{a}_{-1})/\sqrt{2}$. Dashed circle radius denotes the QPN of the polar state. \textbf{a2}. With $\hat{a}_+$ mode eliminated, the state reduces to a single-mode squeezed state that remains squeezed along the $Q_{xz}$ axis while exhibiting QPN-limited fluctuation along the $Q_{yz}$ axis. \textbf{b}. Temporal evolution of the Bell witness evaluated at the optimal orientation angle. Violation of the qubit bound (dash-dotted line) of Bell inequality for two-mode squeezed state (blue circles) is observed over an intermediate range of spin-mixing time, with a minimum value of $W=0.52(7)$ (highlighted by the red diamond) certifying that $48\%$ of the atoms occupy genuine qutrit state. By contrast, the single-mode squeezed state (green squares) yields no violation of the Bell inequality (dashed line). The gradient shading encodes the fraction of certified qutrit particles, spanning from $0\%$ to $100\%$. \textbf{c-d}. Temporal evolution of the linear and nonlinear term of the Bell witness $W_{\rm lin}(t)$ and $W_{\rm non}(t)$, respectively.}
    \label{fig:3}
\end{figure}

Over the past decade, spin-mixing dynamics over long evolution time has proven instrumental in generating metrologically useful non-Gaussian entangled states~\cite{Luo17deterministic,Lucke11twin}, and in realizing quantum phase amplification via time reversal~\cite{Linnemann16Quantum} or quasi-cyclic dynamics~\cite{liu2022nonlinear}. 
In the short-time regime, wherein the $\ket{0}$ mode remains essentially undepleted, spin-mixing dynamics generates spin-nematic squeezing, equivalently known as a two-mode squeezed vacuum~\cite{two_spin_squeezing,Hamley2012,mao2023quantum}. As shown in Fig.~\ref{fig:2}\textbf{c}, this type of state manifests anisotropic Husimi distributions across the spin-nematic planes $ Q_{yz}$-$L_x$ and $ Q_{xz}$-$ L_y$, while retaining full isotropy in both the spin plane ${L}_{x}$-${L}_{y}$ and the nematic plane ${Q}_{xz}$-${Q}_{yz}$. Rotations within the spin-nematic planes are accessible by imprinting spinor phases via tuning the QZS~\cite{SupplementalMaterial}, which concurrently induces contraction or dilation in the other two planes, as depicted in Fig.~\ref{fig:2}\textbf{d}. Crucially, simultaneous squeezing in the two spin-nematic planes is essential for revealing genuine three-level Bell correlations in this work.

\textit{Observation of Bell correlations}---In our experiment, we prepare BECs of $N=3.1\times10^4$ $^{87}$Rb atoms in a crossed optical dipole trap, with all atoms initialized in the $\ket{0}$ mode under a bias magnetic field of $0.99$~G. To observe the violation of Bell inequality, spin-nematic squeezed states are prepared by quenching $q$ to $|c_2|=2\pi\times3.55$~Hz with a microwave dressing field blue-detuned by 7~MHz from the $|1,0\rangle\leftrightarrow|2,0\rangle$ transition. The anisotropic quantum fluctuation of the resulting state can be characterized by the variance of the quadrature operator $\hat{Q}(\theta)=\cos\theta\hat{Q}_{yz}+\sin\theta\hat{L}_x$ and $\hat{Q}'(\theta)=\cos\theta\hat{Q}_{xz}-\sin\theta\hat{L}_y$ in the corresponding spin-nematic planes, where $\theta$ denotes the orientation angle. 
For the initial polar state, the symmetry condition $\langle\hat{Q}(\theta)^2\rangle=\langle\hat{Q}'(\theta)^2\rangle$ holds, and we have 
${W}_{\rm non}(\theta)=6\langle\hat{Q}(\theta)^2\rangle /N$
, where ${W}_{\rm non}(\theta)=\langle\exp(-i\hat{N}_0\theta)\hat{W}_{\rm non}\exp(i\hat{N}_0\theta)\rangle$. These relations are preserved throughout the ideal spin-mixing process and remains robust against particle loss and detection imperfections in our experimental settings. The operator $\hat{Q}(\theta)$ is accessible by first mapping it onto $\hat L_z=\hat{N}_1-\hat{N}_{-1}$ via a sequence of microwave and RF rotations \cite{SupplementalMaterial}, followed by spin-resolved, low-noise absorption imaging~\cite{Luo17deterministic} to extract the populations in the $\ket{\pm1}$ modes. 

Figure~\ref{fig:2}\textbf{e} shows the measured nonlinear Bell witness term $W_{\rm non}(\theta)$ for both the spin-nematic squeezed state generated after $t=50$ ms spin mixing (blue circles) and the initial polar state (white squares) as a function of $\theta$. The squeezed state attains a minimum $W_{\rm non}(\theta_{\rm opt})=0.9(2)$ at $\theta_{\rm opt}=0.79$ rad, corresponding to a spin-nematic squeezing parameter of $\xi^2(\theta_{\rm opt})=-8.2(8)$ dB. Here, we define $\xi^2(\theta)\equiv10\log_{10}(\Delta^2\hat{Q}(\theta)/N)$, which compares quantum noise to the classical shot-noise statistics~\cite{Hamley2012}.
By contrast, the unentangled polar state shows isotropic quantum noise at the shot-noise level, with $W_{\rm non}(\theta) \simeq 6$ uniformly across all angles. 
The linear Bell witness term, on the other hand, is directly related to the fractional population of the $\ket{0}$ mode via $W_{\rm lin}(\theta)=5(1-\langle\hat{N}_0\rangle/N)$.
For spin-nematic squeezed states, $\langle\hat{N}_0\rangle/N \approx 1$, resulting in $W_{\rm lin}(\theta) \approx 0$, consistent with the experimental results presented in Fig.~\ref{fig:2}\textbf{f}.
The full Bell witness $W=W_{\rm lin}+W_{\rm non}$ (Fig.~\ref{fig:2}\textbf{g}) reveals a minimum value of $0.9(2)$, well below the classical bound of $W=2$ (dashed line) at the optimal squeezing direction, thereby providing clear evidence of Bell correlations in our spin-1 system. 

\textit{Certification of qutrit Bell correlation}---To further demonstrate a violation of Bell inequality beyond the qubit bound, it is essential to suppress $\langle\hat{Q}^2\rangle$ so as to minimize $W$. To this end, we vary the spin-mixing time $t$ and characterize the temporal evolution of $W(t)$. At each $t$, the angle $\theta$ is optimized to ensure simultaneous minimization of variances for both $\hat{Q}_{yz}$ and $\hat{Q}_{xz}$. By defining $\hat{a}_{\pm}=(\hat{a}_1\pm\hat{a}_{-1})/\sqrt{2}$, the nematic operators admit the bosonic representation $\hat{Q}_{xz}=\hat{a}_0^{\dagger}a_{-}+\hat{a}_0\hat{a}_{-}^{\dagger}$ and $\hat{Q}_{yz}=i(\hat{a}_0^{\dagger}a_{+}-\hat{a}_0\hat{a}_{+}^{\dagger})$. Consequently, the simultaneous noise reduction of $\hat{Q}_{yz}$ and $\hat{Q}_{xz}$ is equivalent to two-mode squeezing of the $\hat{a}_{\pm}$ modes, with $\hat{a}_0$ serving as the local oscillator mode. As shown in Fig.~\ref{fig:3}\textbf{b} (blue circles), $W(t)$ decreases monotonically during the early stage of evolution. A minimum value of $W=0.52(7)$ is achieved at $t=95$ ms (red diamond), lying below the qubit bound of $W=1$ (dash-dotted line). According to Eq.~(\ref{qutrit}), it is certified that $48\%$ of the total atoms require a qutrit description. To the best of our knowledge, this constitutes the first experimental observation of high-dimensional Bell correlations in a many-body quantum system. Beyond \(t \gtrsim 100\,\mathrm{ms}\), \(W(t)\) increases as the system evolves beyond the optimal regime, eventually losing its ability to certify Bell correlations. In this regime, the state exhibits pronounced non-Gaussian characteristics~\cite{gerving_non-equilibrium_2012,liu2022nonlinear}, manifested in the growth of both $W_{\rm lin}$ (Fig.~\ref{fig:3}\textbf{c}) arising from the depletion of the $\ket{0}$ mode, and $W_{\rm non}$ (Fig.~\ref{fig:3}\textbf{d}) due to the over-squeezing and wrapping of the state around the Bloch sphere. Such non-Gaussian states can still exhibit strong quantum correlations, the detection of which, however, lies beyond the capability of the Bell witness in Eq.~(\ref{W}) involving only the first and second moments of the nematic operators~\cite{GuoPRL23}.

To further demonstrate that the observed Bell witness violation originates from genuine spin-1 correlations, we implement a three-step sequence to selectively eliminate the $\hat{a}_+$ mode population as a control experiment. First, the $\hat{a}_0$ and $\hat{a}_+$ modes are exchanged via an RF $\pi/2$ pulse, leaving the $\hat{a}_-$ mode unaffected. Second, the transferred population in $|1,0\rangle$ is driven to $|2,0\rangle$ by a microwave $\pi$ pulse and subsequently removed with a resonant probe beam. Finally, a second RF $\pi/2$ pulse swaps $\hat{a}_+$ and $\hat{a}_0$ modes back, leaving the $\hat{a}_+$ mode in the vacuum. This procedure converts the two-mode squeezing into single-mode squeezing of the $\hat a_-$ quadrature (Fig.~\ref{fig:3}\textbf{a1, a2}), essentially reducing the genuine local dimensionality from three to two, even though the atoms continue to occupy all three Zeeman modes. The resulting Bell witness, shown in Fig.~\ref{fig:3}\textbf{b} (green squares), exhibits no violation across the entire range of evolution time, a direct consequence of this dimensionality reduction. We emphasize that the absent violation cannot be attributed to the degradation of the $\hat{a}_-$ mode squeezing: the optimal squeezing parameter at $t=95$ ms increases only marginally from $-11.8(7)$ dB to $-10.3(9)$ dB throughout the sequence. The contrasting Bell witness responses under two-mode and single-mode squeezing establish that access to the full three-dimensional local Hilbert space is essential for surpassing the qubit bound of the Bell witness.

\textit{Conclusions}---In this work, we introduce a multipartite high-dimensional Bell correlation witness that is directly accessible with collective spin and nematic observables and use it to observe genuinely three-level Bell correlations in a spin-1 $^{87}$Rb BEC.
Exploiting spin-mixing dynamics, we generate up to $-11.8(7)$ dB of spin-nematic squeezing in an ensemble of $N\simeq 3.1\times 10^4$ atoms and achieve a Bell witness violation that surpasses the minimum bound attainable by a collection of $N$ qubits, thereby providing evidence of multipartite qutrit Bell correlations with collective measurements.
Our results establish spinor condensates as a scalable platform for exploring high-dimensional Bell correlations in the many-body regime, and introduce a practical route to witness local Hilbert-space dimensionality without single-particle resolution.
Our measurements further highlight that simultaneously squeezed nematic quadratures constitute the essential resource enabling access to spin-1 correlations that go beyond a qubit description.
In the future, extending our framework to capture non-Gaussian correlations, for instance by incorporating higher moments \cite{GuoPRL23}, should enable tighter dimensionality certification and improved robustness to depletion and technical noise.
More broadly, our approach opens new opportunities for the validation of complex many-body resources going beyond qubit correlations, complementing the capabilities already demonstrated in multiparameter quantum sensing~\cite{cao_joint_2025} and opening avenues for randomness generation, while sharpening the boundary between effective two-level descriptions and the richer nonlocal structures of interacting multilevel quantum systems.

\textit{Acknowledgments}---We thank J. H. Cao, Z. P. Yue, M. K. Tey, Z. X. Hua and Y. Q. Zou for helpful discussions. This work is supported by the National
Natural Science Foundation of China (NSFC) (Grants No. 12361131576 and No. 92265205, 92476205, and 92565306), and by
the Innovation Program for Quantum Science and Technology (2021ZD0302104). GMR acknowledges financial support by the European Union under ERC Advanced Grant TAtypic, Project No. 101142236. MF was supported by the Swiss National Science Foundation Ambizione Grant No. 208886, and by The BrancoWeiss Fellowship – Society in Science, administered by the ETH Z\"urich.

\let\oldaddcontentsline\addcontentsline
\renewcommand{\addcontentsline}[3]{}
%

\let\addcontentsline\oldaddcontentsline

\clearpage
\newpage
\noindent

\renewcommand{\thetable}{S\arabic{table}}  
\renewcommand{\thepage}{S\arabic{page}}  
\renewcommand{\thefigure}{S\arabic{figure}}
\renewcommand{\theHfigure}{S\arabic{figure}}
\renewcommand{\theequation}{S\arabic{equation}}
\setcounter{page}{1}
\setcounter{figure}{0}
\setcounter{table}{0}
\setcounter{section}{0}
\setcounter{equation}{0}

\widetext

\begin{center}
{\large\bfseries Supplemental Material for\\
``Observing Bell Inequality Violation Beyond the Qubit Bound in a Spinor Bose--Einstein Condensate''}
\end{center}
\normalsize
\begin{center}
Wenxin Xu$^{1,*}$,
Junshuang Hu$^{1,*}$,
Guillem M\"{u}ller-Rigat$^{2}$,
Xinwei Li$^{3}$,
Qi Liu$^{4, \dagger}$,
Matteo Fadel$^{5,\ddagger}$,
and Li You$^{1,3,6,7,\S}$
\end{center}

\begin{center}
{\small\textit{
$^{1}$State Key Laboratory of Low Dimensional Quantum Physics, Department of Physics, Tsinghua University, Beijing 100084, China \\[4pt]
$^{2}$Faculty of Physics, Astronomy and Applied Computer Science, Jagiellonian University, ul. {\L}ojasiewicza 11, 30-348 Krak\'{o}w, Poland \\[4pt]
$^{3}$Beijing Academy of Quantum Information Sciences, Beijing 100193, China \\[4pt]
$^{4}$National Laboratory of Solid State Microstructures and School of Physics, Collaborative Innovation Center of Advanced Microstructures, Nanjing University, Nanjing 210093, China \\[4pt]
$^{5}$Department of Physics, ETH Z\"{u}rich, 8093 Z\"{u}rich, Switzerland \\[4pt]
$^{6}$Frontier Science Center for Quantum Information, Beijing, China \\[4pt]
$^{7}$Hefei National Laboratory, Hefei, Anhui 230088, China \\[4pt]}
$^{\ast}$These authors contributed equally to this work. \\[4pt]
}
\end{center}
\suppressfloats

\tableofcontents

\clearpage
\newpage

\section{Introduction to spin-nematic squeezing}

In a spin-$1$ spinor Bose--Einstein condensate confined to a single spatial mode (single-mode approximation), the second-quantized field operators factorize as $\hat\Psi_m(\mathbf r)=\phi(\mathbf r)\hat a_m$ for Zeeman sublevels $m\in\{-1,0,+1\}$, with $[\hat a_m,\hat a_{m'}^\dagger]=\delta_{m{m'}}$ and fixed total atom number $\hat N=\sum_m \hat a_m^\dagger \hat a_m\simeq N$. For contact interactions, the most general spin-rotation-invariant two-body Hamiltonian is
\[
\hat H_{\rm int}=\frac{1}{2}\!\int\! d^3r\;\Big[c_0 \hat n^2(\mathbf r) + c_2 \hat{\mathbf F}^2(\mathbf r)\Big].
\]
The coupling constants are $c_0=\frac{4\pi\hbar^2}{3M}(a_0+2a_2)$ and $c_2=\frac{4\pi\hbar^2}{3M}(a_2-a_0)$ in terms of $s$-wave scattering lengths $a_F$ in total-spin channels $F=0,2$, where $M$ is the atom's mass. 
Furthermore, $\hat n=\sum_m \hat\Psi_m^\dagger\hat\Psi_m$ and $\hat{\mathbf F}=\sum_{mn}\hat\Psi_m^\dagger\,\mathbf f_{mn}\,\hat\Psi_n$ with $\mathbf f=(f_x,f_y,f_z)$ the spin-$1$ matrices given by
\begin{align}
f_x &= \frac{1}{\sqrt{2}}
\begin{pmatrix}
0 & 1 & 0\\
1 & 0 & 1\\
0 & 1 & 0
\end{pmatrix}, \qquad
f_y = \frac{1}{\sqrt{2}}
\begin{pmatrix}
0 & -i & 0\\
i & 0 & -i\\
0 & i & 0
\end{pmatrix}, \qquad
f_z =
\begin{pmatrix}
1 & 0 & 0\\
0 & 0 & 0\\
0 & 0 & -1
\end{pmatrix}. \label{eq:fxyz}
\end{align}
They satisfy the angular momentum algebra $[f_\alpha,f_\beta]=i\epsilon_{\alpha\beta\gamma}f_\gamma$ for $\alpha,\beta,\gamma\in\{x,y,z\}$, as well as the spin-1 identities $f_x^2+f_y^2+f_z^2 = 2\,\mathbb I_3$ and $\mathrm{Tr}(f_\alpha)=0$, where $\mathbb I_3$ is the $3\times3$ identity matrix.
The spin-1 matrices allow us to define the collective spin operators in terms of the bosonic operators as (Jordan–Schwinger map)
\begin{equation}
\hat{\mathbf L} =
\sum_{m,m'=-1}^{+1}
\hat a_m^\dagger \, \mathbf f_{m m'} \, \hat a_{m'},
\end{equation}
giving explicitly
\begin{subequations}\label{suppeq:Ldefs}
\begin{align}
\hat L_x &=
\frac{1}{\sqrt{2}}
\left(
\hat a_{+1}^\dagger \hat a_0 + \hat a_0^\dagger \hat a_{+1} + \hat a_0^\dagger \hat a_{-1} + \hat a_{-1}^\dagger \hat a_0
\right), \\
\hat L_y &=
\frac{i}{\sqrt{2}}
\left(
-\hat a_{+1}^\dagger \hat a_0 + \hat a_0^\dagger \hat a_{+1}
-\hat a_0^\dagger \hat a_{-1} + \hat a_{-1}^\dagger \hat a_0
\right), \\
\hat L_z &=
\hat a_{+1}^\dagger \hat a_{+1} - \hat a_{-1}^\dagger \hat a_{-1} = \hat N_{+1} - \hat N_{-1},
\end{align}
\end{subequations}
where $\hat N_m = \hat a_m^\dagger \hat a_m$ giving $\hat N = \hat N_{+1} + \hat N_0 + \hat N_{-1}$, and satisfying the angular momentum algebra $[\hat L_\alpha, \hat L_\beta] = i \epsilon_{\alpha\beta\gamma} \hat L_\gamma$.
Similarly, we introduce the nematic (quadrupole) tensor, defined as the second-quantized sum of single-particle spin-1 quadrupoles, as
\begin{equation}
\hat Q_{\alpha\beta} = \sum_{m,{m'}=-1}^{+1}
\hat a_m^\dagger
\left[
(f_\alpha f_\beta + f_\beta f_\alpha)
-\frac{4}{3}\,\delta_{\alpha\beta}\,\mathbb I_3
\right]_{m{m'}}
\hat a_{m'} .
\label{eq:Qdef}
\end{equation} 
By construction, $\hat Q_{\alpha\beta}$ is symmetric, $\hat Q_{\alpha\beta}=\hat Q_{\beta\alpha}$, traceless, $\sum_\alpha \hat Q_{\alpha\alpha}=0$, and transforms as a rank-two tensor under rotations.
In particular, we have $\hat Q_{zz} = \frac{2}{3}
\big(
\hat a_{+1}^\dagger \hat a_{+1} + \hat a_{-1}^\dagger \hat a_{-1}
\big) - \frac{4}{3}
\hat a_0^\dagger \hat a_0 = \frac{2}{3}\hat N - 2 \hat N_0$.

Under the single-mode approximation one obtains $\hat H_{\rm int}=\frac{c_0\Lambda}{2}\hat N(\hat N-1)+\frac{c_2\Lambda}{2}\hat{\mathbf L}^2$ with $\Lambda=\int d^3r\,|\phi(\mathbf r)|^4$ and the collective spin operator $\hat{\mathbf L}^2 = \hat L_x^2 + \hat L_y^2 + \hat L_z^2$. 
For fixed total $N$ the spin-independent term is a constant and can be dropped. 
Including the quadratic Zeeman shift from a bias field, $\hat H_{q}=-q\,\hat N_0$ with $\hat N_0=\hat a_0^\dagger\hat a_0$ (up to an additive constant and after removing the linear Zeeman term by working in a rotating frame at fixed magnetization), one arrives at the effective Hamiltonian
\begin{equation}\label{supp:theH}
\hat H=\frac{c_2}{2N}\,\hat{\mathbf L}^{\,2}-q\,\hat N_0,
\end{equation}
where $c_2$ denotes the appropriately rescaled spin-dependent interaction strength (absorbing $\Lambda$ and the $N$-dependent normalization).
Using the definitions in Eqs.~\eqref{suppeq:Ldefs} we obtain
\begin{equation}
    \hat{H} = \frac{c_2}{2N} \Big[
2(\hat{a}_1^\dagger \hat{a}_{-1}^\dagger \hat{a}_0 \hat{a}_0 + \text{h.c.}) \\
+ (2\hat{N}_0 - 1)(N - \hat{N}_0)
\Big] - q\hat{N}_0 \;,
\end{equation}
which is Eq.~\eqref{Ham} of the main text.

\subsection{The polar Fock state}

Initially, the system is initialized in the polar Fock state
\begin{equation}
\lvert \mathrm{polar}\rangle
= \frac{(\hat a_0^\dagger)^N}{\sqrt{N!}} \lvert \mathrm{vac}\rangle,
\qquad
\langle \hat N_0 \rangle = N,\quad \langle \hat N_{+1} \rangle = \langle \hat N_{-1} \rangle = 0 .
\end{equation}
For this state, all first moments of the collective spin components vanish,
\begin{equation}
\langle \hat L_x \rangle
= \langle \hat L_y \rangle
= \langle \hat L_z \rangle
= 0 .
\end{equation}
Second moments of the collective spin components are
\begin{equation}
\langle \hat L_x^2 \rangle
= \langle \hat L_y^2 \rangle
= N,
\qquad
\langle \hat L_z^2 \rangle
= 0 ,
\end{equation}
while all mixed second moments vanish:
\begin{equation}
\langle \hat L_x \hat L_y + \hat L_y \hat L_x \rangle = 0,
\qquad
\langle \hat L_x \hat L_z + \hat L_z \hat L_x \rangle = 0,
\qquad
\langle \hat L_y \hat L_z + \hat L_z \hat L_y \rangle = 0 .
\end{equation}
Hence the variances are
\begin{equation}
(\Delta \hat{L}_x)^2 = (\Delta \hat{L}_y)^2 = N,
\qquad
(\Delta \hat{L}_z)^2 = 0 .
\end{equation}
Using the definition $\hat Q_{zz} = \frac{2}{3}\hat N - 2\hat N_0$,
the expectation value of quadrupole (nematic) operators are
\begin{equation}
\langle \hat Q_{zz} \rangle
= -\frac{4N}{3},
\end{equation}
and by rotational symmetry about the $z$ axis,
\begin{equation}
\langle \hat Q_{xx} \rangle
= \langle \hat Q_{yy} \rangle
= \frac{2N}{3},
\qquad
\langle \hat Q_{xy} \rangle
= \langle \hat Q_{xz} \rangle
= \langle \hat Q_{yz} \rangle
= 0 
\qquad
\langle \hat Q_{xy}^2 \rangle = 0
\qquad
\langle \hat Q_{xz}^2 \rangle
= \langle \hat Q_{yz}^2 \rangle
= N .
\end{equation}

For $q\in(0, 2|c_2|)$, the system initialized at the polar state becomes dynamically unstable, leading to the generation of atomic pairs in $\left|\pm1\right\rangle$ from $\left|0\right\rangle$ and vice versa. Since the evolution occurs within the decoherence-free subspace at $\hat{L}_z=0$, the quantum coherence is protected from magnetic field fluctuation to first order.

\subsection{Spin-nematic squeezing}

For short evolution times, spin-exchange collisions
\(\ket{0}\ket{0}\rightarrow \ket{+1}\ket{-1}\) create only a small population in \(m=\pm1\).
We therefore linearize the dynamics around the polar ``pump field'' by replacing
\(\hat a_0\simeq \sqrt{N}\) and keeping terms up to quadratic order
in \(\hat a_{\pm1}\).

Using
\begin{equation}
-q\hat N_0 = -q(N-\hat N_{+1}-\hat N_{-1})
= q(\hat N_{+1}+\hat N_{-1}) + \text{const},
\end{equation}
and expanding \(\hat{\mathbf L}^2\) about the polar state, the Hamiltonian \eqref{supp:theH} reduces (up to constants) to
\begin{equation}
\hat H_{\mathrm{quad}}
=
(q+c_2)\left(\hat N_{+1}+\hat N_{-1}\right)
+
c_2\left(
\hat a_{+1}^\dagger \hat a_{-1}^\dagger
+
\hat a_{+1}\hat a_{-1}
\right).
\label{eq:Hquad}
\end{equation}
This is the standard parametric-amplifier Hamiltonian describing
spin-mixing dynamics.

Let us introduce the symmetric and antisymmetric modes by the linear combinations
\begin{equation}
\hat a_+=\frac{\hat a_{+1}+\hat a_{-1}}{\sqrt{2}},
\qquad
\hat a_-=\frac{\hat a_{+1}-\hat a_{-1}}{\sqrt{2}} .
\end{equation}
In this basis,
\begin{equation}
\hat a_{+1}^\dagger\hat a_{-1}^\dagger
=
\frac{(\hat a_+^\dagger)^2-(\hat a_-^\dagger)^2}{2},
\qquad
\hat N_{+1}+\hat N_{-1}=
\hat a_+^\dagger\hat a_+ + \hat a_-^\dagger\hat a_- ,
\end{equation}
and \eqref{eq:Hquad} becomes
\begin{equation}
\hat H_{\mathrm{quad}}
=
(q+c_2)\!\left(\hat a_+^\dagger\hat a_+ + \hat a_-^\dagger\hat a_- \right)
+
\frac{c_2}{2}\!\left[(\hat a_+^\dagger)^2+\hat a_+^2\right]
-
\frac{c_2}{2}\!\left[(\hat a_-^\dagger)^2+\hat a_-^2\right],
\label{eq:Hsa}
\end{equation}
showing that the short-time dynamics results in a two-mode squeezed vacuum with independent single-mode squeezing in the symmetric and antisymmetric modes.

The linearized Heisenberg equations for either mode \(\mu\in\{+,-\}\) are
\begin{equation}
\frac{d}{dt}\hat a_\mu
=
-i\left[(q+c_2)\hat a_\mu + \kappa_\mu \hat a_\mu^\dagger\right],
\qquad
\kappa_+=+c_2,\;\; \kappa_-=-c_2 .
\end{equation} 
The eigenvalues of the dynamical matrix describing the time evolution of $\{\hat a_\mu, \hat a_\mu^\dagger\}$ give the characteristic frequency
\begin{equation}
\omega^2 = q(q+2c_2).
\label{eq:omega}
\end{equation}

To leading order in small excitations, the transverse spin and nematic operators reduce to quadratures of \(\hat a_+\) and \(\hat a_-\). This allows us to write the spin--nematic operators near the polar state as
\begin{align}
\hat L_x &\simeq \sqrt{2N}\,\hat X_+,
&
\hat Q_{yz} &\simeq -\sqrt{2N}\,\hat P_+,
\\
\hat Q_{xz} &\simeq \sqrt{2N}\,\hat X_-,
&
\hat L_y &\simeq -\sqrt{2N}\,\hat P_-,
\end{align}
where
\(
\hat X_\mu=(\hat a_\mu+\hat a_\mu^\dagger)/\sqrt{2}
\)
and
\(
\hat P_\mu=(\hat a_\mu-\hat a_\mu^\dagger)/(i\sqrt{2})
\).

The Heisenberg equations form a closed system within two independent subspaces:
\begin{equation}
\frac{d}{dt}
\begin{pmatrix}
\hat L_x \\[2pt]
\hat Q_{yz}
\end{pmatrix}
=
\begin{pmatrix}
0 & -q \\
(q+2c_2) & 0
\end{pmatrix}
\begin{pmatrix}
\hat L_x \\[2pt]
\hat Q_{yz}
\end{pmatrix},
\label{eq:block1}
\end{equation}
and
\begin{equation}
\frac{d}{dt}
\begin{pmatrix}
\hat L_y \\[2pt]
\hat Q_{xz}
\end{pmatrix}
=
\begin{pmatrix}
0 & q \\
-(q+2c_2) & 0
\end{pmatrix}
\begin{pmatrix}
\hat L_y \\[2pt]
\hat Q_{xz}
\end{pmatrix}.
\label{eq:block2}
\end{equation}

For \(\omega^2>0\), the solutions are oscillatory, while for the instability regime $\omega^2<0$ become exponentially increasing.

\subsection*{Instability and the squeezing regime}

Spin-nematic squeezing corresponds to exponential amplification of one quadrature
and attenuation of its conjugate. This occurs when the frequency
\eqref{eq:omega} becomes imaginary:
\begin{equation}
\omega^2 = q(q+2c_2) < 0.
\end{equation}
Equivalently,
\begin{equation}
\min(0,-2c_2) < q < \max(0,-2c_2).
\end{equation}

For ferromagnetic interactions \(c_2<0\), this reduces to
\begin{equation}
0 < q < 2|c_2|,
\end{equation}
which is the instability window quoted in the main text.
In this regime the solutions of \eqref{eq:block1}--\eqref{eq:block2}
become hyperbolic, leading to exponential growth and
spin-nematic squeezing in appropriate rotated quadratures of
\((\hat L_x,\hat Q_{yz})\) and \((\hat L_y,\hat Q_{xz})\).

Concretely, since in the instability window the characteristic frequency is imaginary we thus write
\begin{equation}
\omega = i\gamma,
\qquad
\gamma \equiv \sqrt{-q(q+2c_2)} > 0 .
\end{equation}
Then the closed equations~\eqref{eq:block1}
yield the hyperbolic solutions
\begin{align}
\hat L_x(t)
&=
\hat L_x(0)\cosh(\gamma t)
-
\frac{q}{\gamma}\,\hat Q_{yz}(0)\sinh(\gamma t),
\label{eq:Lx_hyp}
\\
\hat Q_{yz}(t)
&=
\hat Q_{yz}(0)\cosh(\gamma t)
+
\frac{q+2c_2}{\gamma}\,\hat L_x(0)\sinh(\gamma t).
\label{eq:Qyz_hyp}
\end{align}
Similarly, for the second pair one finds
\begin{align}
\hat L_y(t)
&=
\hat L_y(0)\cosh(\gamma t)
+
\frac{q}{\gamma}\,\hat Q_{xz}(0)\sinh(\gamma t),
\label{eq:Ly_hyp}
\\
\hat Q_{xz}(t)
&=
\hat Q_{xz}(0)\cosh(\gamma t)
-
\frac{q+2c_2}{\gamma}\,\hat L_y(0)\sinh(\gamma t).
\label{eq:Qxz_hyp}
\end{align}
For short times,
\begin{align}
\hat L_x(t) &= \hat L_x(0) - q\,t\,\hat Q_{yz}(0) + \mathcal O(t^2),\\
\hat Q_{yz}(t) &= \hat Q_{yz}(0) + (q+2c_2)\,t\,\hat L_x(0) + \mathcal O(t^2),
\end{align}
(and analogously for \(\hat L_y,\hat Q_{xz}\)).

Define now rotated quadratures in the \((\hat L_x,\hat Q_{yz})\) subspace by
\begin{equation}
\hat X_\theta \equiv \hat L_x\cos\theta + \hat Q_{yz}\sin\theta.
\end{equation}
In the polar initial state, the relevant second moments at \(t=0\) are
\begin{equation}
\langle \hat L_x\rangle=\langle \hat Q_{yz}\rangle=0,\qquad
(\Delta \hat{L}_x)^2_0 = N,\qquad
(\Delta \hat{Q}_{yz})^2_0 = N,\qquad
\mathrm{Cov}_0(\hat L_x,\hat Q_{yz})=0,
\end{equation}
where \(\mathrm{Cov}(\hat A,\hat B)=\tfrac12\langle \hat A\hat B+\hat B\hat A\rangle-\langle\hat A\rangle\langle\hat B\rangle\).

Using \eqref{eq:Lx_hyp}--\eqref{eq:Qyz_hyp} and keeping the leading nontrivial
terms in \(t\), the variances evolve as
\begin{align}
(\Delta \hat{L}_x)^2(t)
&=
N\Big[1+q^2 t^2\Big]+\mathcal O(t^3),
\label{eq:VarLx_short}
\\
(\Delta \hat{Q}_{yz})^2(t)
&=
N\Big[1+(q+2c_2)^2 t^2\Big]+\mathcal O(t^3),
\label{eq:VarQyz_short}
\\
\mathrm{Cov}(\hat L_x,\hat Q_{yz})(t)
&=
2 c_2 N t+\mathcal O(t^2).
\label{eq:Cov_short}
\end{align}
Therefore the variance of the rotated quadrature \(\hat X_\theta\) is
\begin{align}
(\Delta \hat{X}_\theta)^2(t)
&=
(\Delta \hat{L}_x)^2(t)\cos^2\theta
+
(\Delta \hat{Q}_{yz})^2(t)\sin^2\theta
+
2\,\mathrm{Cov}(\hat L_x,\hat Q_{yz})(t)\sin\theta\cos\theta
\nonumber\\
&=
N\Big[1+2c_2\,t\,\sin(2\theta)\Big]+\mathcal O(t^2).
\label{eq:VarXtheta_short}
\end{align}
The optimal angles for short-time evolution are \(\theta=\pm \pi/4\), for which
\begin{equation}
(\Delta \hat{X}_{\pm \pi/4})^2(t)
=
N\Big[1\pm 2c_2\,t\Big]+\mathcal O(t^2).
\label{eq:sq_asq_short}
\end{equation}
Hence one quadrature is squeezed below the polar reference level \(N\),
while the orthogonal quadrature is anti-squeezed by the same amount to leading order.
(If \(q+c_2>0\), the squeezed quadrature is \(\theta=+\pi/4\); if \(q+c_2<0\), it is \(\theta=-\pi/4\).)

The same analysis applies to the \((\hat L_y,\hat Q_{xz})\) subspace, yielding an
independent pair of squeezed/anti-squeezed quadratures with identical short-time scaling.

\subsection{Numerical simulations with the truncated Wigner method}
The spin-mixing dynamics for spin-1 condensates under the single-mode approximation is described by the Hamiltonian
\begin{equation}
\hat{H} = \frac{c_2}{2N} \Big[2(\hat{a}_1^\dagger \hat{a}_{-1}^\dagger \hat{a}_0 \hat{a}_0 + \text{h.c.}) + (2\hat{N}_0 - 1)(N - \hat{N}_0)
\Big] - q\hat{N}_0 .
\end{equation}
To account for the effect of noise, which in our experiment is mostly due to atom losses, it is necessary to solve the master equation
\begin{equation}
    \frac{d{\hat{\rho}}}{dt}=-i[{\hat{H}}(t),{\hat{\rho}}]+\sum_{i=0,\pm1}\gamma {\cal D}[{\hat{a}}_i]{\hat{\rho}},
\label{master}
\end{equation}
with ${\cal D}[{\hat{a}}_i]{\hat{\rho}}={\hat{a}}_i{\hat{\rho}}{\hat{a}}_i^{\dagger}-\{{\hat{a}}_i^{\dagger}{\hat{a}}_i,{\hat{\rho}}\}/2$ and $\gamma$ the single-particle loss rate.

One efficient approach to solve this equation is the so-called truncated Wigner method~\cite{Blakie:2008aa}, which describes the dynamics through the stochastic differential equations (SDEs)~\cite{liu2022nonlinear}:
\begin{equation}
			\left\{
			\begin{aligned}\label{rfsde}
				d\psi_1 &=-ic_2'\left[\psi_0^2\psi_{-1}^*+(|\psi_1|^2-|\psi_{-1}|^2+|\psi_0|^2)\psi_1\right]dt-\frac{\gamma}{2}\psi_1dt+\sqrt{\frac{\gamma}{2}}d\xi_1(t), \\
				d\psi_0 &=-ic_2'\left[2\psi_1\psi_{-1}\psi_0^*+\left(|\psi_1|^2+|\psi_{-1}|^2\right)\psi_0\right]dt+i\psi_0\left[qdt+\eta_qd\xi_q(t)\right]-\frac{\gamma}{2}\psi_0dt+\sqrt{\frac{\gamma}{2}}d\xi_0(t),\\
				d\psi_{-1} &=-ic_2'\left[\psi_0^2\psi_1^*+(|\psi_{-1}|^2-|\psi_1|^2+|\psi_0|^2)\psi_{-1}\right]dt-\frac{\gamma}{2}\psi_{-1}dt+\sqrt{\frac{\gamma}{2}}d\xi_{-1}(t).
			\end{aligned}
			\right.
		\end{equation}
Here, $c_2^\prime={c_2(t)}/{N(t)}$ denotes the drifting spin-exchange strength due to atom loss, and $d\xi_i(t)$ the complex Wiener noise increments satisfying $\overline{d\xi_i(t)}=0$, $\overline{d\xi_i^*(t)d\xi_j(t)}=\delta_{i,j}dt$. The contribution of quantum noise is incorporated into the probability distribution of the Wigner function for the initial polar state, and can be sampled according to
\begin{equation}
\begin{aligned}
&\psi_{\pm1}=\dfrac{1}{2}(\alpha_{\pm1}+i\beta_{\pm1}),\\
&\psi_0=\sqrt{N}+\dfrac{1}{2}(\alpha_{0}+i\beta_{0}),
\end{aligned}
\end{equation}
where $\alpha_i$ and $\beta_i$ are independent real numbers following standard normal distribution. We take $10^4$ samplings in our simulations, each of which is associated with an evolution trajectory. 
The average and variance of mode population are then obtained from the average of simulated trajectories, calculated according to $\langle{\hat N}_i\rangle=\overline{\psi_i^*\psi_i}-1/2$ and $(\Delta \hat {N}_i)^2=(\Delta\psi_i^*\psi_i)^2-1/4$.

The parameters used in the simulations of spin-mixing dynamics include: initial atom number $N=31000$, $c_2=-2\pi\times3.55$ Hz, $q=2\pi\times3.55$ Hz and loss rate $\gamma=0.069~\rm{s^{-1}}$. Extra detection noise of $(\Delta \hat{L}_z)_{\rm det}=23.1$ is taken into account through a Gaussian convolution of the Wigner distribution. Figure 3 in the main text also considers a $2\%$ calibration error of the rotation angle for aligning the squeezing axis along $\hat{Q}_{yz/xz}$.

\clearpage
\newpage

\section{Bell inequality and dimension witness}

\subsection{Bell inequality and operator}

In the main text we have introduced the Bell inequality~\eqref{eq:5dimPIBI}
\begin{equation}\label{supp:BellIn}
    B =\mathcal{P}_{-1|0}+\mathcal{P}_{-1|1} + \mathcal{P}_{1|0}+\mathcal{P}_{1|1} +\mathcal{P}_{-1-1|00} + \mathcal{P}_{-1-1|11} + \mathcal{P}_{11|00} + \mathcal{P}_{11|11} - 2\mathcal{P}_{-11|01} - 2\mathcal{P}_{-11|10} \geq 0.
\end{equation}
Within the framework of quantum mechanics, this quantity can be interpreted as the expectation value of an operator $\hat B$, meaning ${B}=\langle\hat B\rangle$.
For this, we assume the probability distributions to originate according to Born's rule from measurements of a many-body quantum state $\hat \rho$.
In general, these measurements are positive operator-valued measures (POVMs) acting on the local Hilbert space $\mathcal{H}$ of each party.
Let us define the POVMs operator associated to measurement $\chi\in\{0,1\}$ and result $a\in\{-1,0,1\}$ as
$\hat \pi_{a|\chi}$.
We introduce the collective one- and two-body probabilities
\begin{equation}
    \mathcal{P}_{a|\chi} = \sum_{i} \left\langle \hat{\pi}^{(i)}_{a|\chi} \right\rangle = \langle \hat{\Pi}_{a|\chi} \rangle  \;,\qquad
    \mathcal{P}_{ab|\chi\gamma} =  \sum_{i\neq j} \left\langle  \hat{\pi}^{(i)}_{a|\chi} \hat{\pi}^{(j)}_{b|\gamma} \right\rangle = \left\langle \dfrac{1}{2} \{\hat{\Pi}_{a|\chi},\hat{\Pi}_{b|\gamma}\}  -\hat{\Pi}_{ab|\chi\gamma}  \right\rangle \ ,
\end{equation}
where we have used the definitions of collective projector operators
\begin{equation}\label{supp:PiDef}
    \hat{\Pi}_{a|\chi} = \sum_{i}\hat{\pi}^{(i)}_{a|\chi} \;,\qquad 
    \hat{\Pi}_{ab|\chi\gamma} = \dfrac{1}{2} \sum_{i}\{\hat{\pi}^{(i)}_{a|\chi},\hat{\pi}^{(i)}_{b|\gamma} \} \;.
\end{equation}
Due to PI symmetry of the Bell inequality~\eqref{supp:BellIn}, the associated Bell operator can be expressed only in terms of collective operators as
\begin{align}
  \hat{{B}} = (\hat{\Pi}_{-1|0}-\hat{\Pi}_{1|1})^2+(\hat{\Pi}_{-1|1}-\hat{\Pi}_{1|0})^2+\hat{{B}}_1\ ,
\end{align}
where $\hat{{B}}_1 = \hat{\Pi}_{-1|0}+\hat{\Pi}_{-1|1}+\hat{\Pi}_{1|0}+\hat{\Pi}_{1|1} 
    -\hat{\Pi}_{-1-1|00} - \hat{\Pi}_{-1-1|11} -\hat{\Pi}_{11|00} - \hat{\Pi}_{11|11} +2(\hat{\Pi}_{-11|01} +\hat{\Pi}_{-11|10})$.
For projective measurements at fixed $\chi$, we have $\hat\pi_{a|\chi}\hat\pi_{b|\chi}=\delta_{ab}\hat\pi_{b|\chi}$, giving $\hat{\Pi}_{ab|xx}=\delta_{ab}\hat{\Pi}_{a|x}$, and the one-body term further simplifies to
\begin{equation}\label{supp:B1op}
    \hat{{B}}_1 = 2(\hat{\Pi}_{-11|01} + \hat{\Pi}_{-11|10}) \ .
\end{equation}
This give us the Bell operator
\begin{equation}
    \hat{B} = (\hat{\Pi}_{-1|0}-\hat{\Pi}_{1|1})^2 +(\hat{\Pi}_{-1|1}-\hat{\Pi}_{1|0})^2 + 2(\hat\Pi_{-11|01}+\hat\Pi_{-11|10}).
    \label{supp:BellOpPi}
\end{equation}

\subsection{Dimension bounds}

As we will show here, the Bell operator Eq.~\eqref{supp:BellOpPi} allows us to construct a Bell dimension witness, meaning an inequality that can be violated only by Bell correlations between systems with local Hilbert space dimensionality $\dim \mathcal{H} \geq d$.
The derivation of such dimensionality witnesses is in general an extremely demanding task, especially in the multipartite setting.
However, the structure of the Bell operator Eq.~\eqref{supp:BellOpPi} enables a scalable relaxation of this problem. 

Note that the first two terms in Eq.~\eqref{supp:BellOpPi} are positive semidefinite, meaning $\langle (\hat{\Pi}_{-1|0}-\hat{\Pi}_{1|1})^2+(\hat{\Pi}_{-1|1}-\hat{\Pi}_{1|0})^2 \rangle\geq 0$.
Therefore, a violation of the associated Bell inequality depends only on the negativity of the term $\langle \hat{{B}}_1 \rangle$.
Looking at its definition Eq.~\eqref{supp:B1op}, we see that this operator can be written as the linear combination of one-body terms
\begin{align}
    \hat{{B}}_1 &= 2(\hat{\Pi}_{-11|01} + \hat{\Pi}_{-11|10}) \\
    &= \sum_{i}\{\hat{\pi}^{(i)}_{-1|0},\hat{\pi}^{(i)}_{1|1} \} + \sum_{i}\{\hat{\pi}^{(i)}_{-1|1},\hat{\pi}^{(i)}_{1|0} \} \\
    &= \sum_i \beta^{(i)} \;,
\end{align}
where we have introduced (omitting superscript $(i)$ on each term and using $\hat\pi_{a|\chi}\hat\pi_{b|\chi}=\delta_{ab}\hat\pi_{b|\chi}$)
\begin{equation}
    \hat{\beta} = \{\hat{\pi}_{-1|0},\hat{\pi}_{1|1} \} + \{\hat{\pi}_{-1|1},\hat{\pi}_{1|0} \} = \hat{\pi}_{-1|0} + \hat{\pi}_{1|1} - (\hat{\pi}_{-1|0} - \hat{\pi}_{1|1})^2 +\hat{\pi}_{-1|1} + \hat{\pi}_{1|0} - (\hat{\pi}_{-1|1} - \hat{\pi}_{1|0})^2  .
    \label{eq:Localterm}
\end{equation}
This leaves us with the task of bounding $\hat \beta$ for a single party.
Indeed, suppose that for a qudit of dimension $d$ the one-body term cannot go below $\beta_d$, meaning $\langle \hat{\beta} \rangle_d \geq  \beta_d$, then
\begin{equation}
   \langle{\hat{B}}\rangle\geq N \beta_d \ ,
    \label{supp:DimBell}
\end{equation}
is a proper Bell dimension witness for arbitrary number of subsystems $N$. 
Being based on a Bell inequality, this criterion is valid regardless of the description of the measurement settings and without relying on trusting their implementation.
A violation of inequality~\eqref{supp:DimBell} implies that the observed statistics require qudits of dimension at least $d+1$ to be reproduced. 

Furthermore, following the same reasoning, linearity of Eq.~\eqref{supp:B1op} allows us to conclude that for a system composed by $n_d\leq N$ parties with dimension less or equal than $d$ we have
\begin{eqnarray}\label{supp:Bdim}
    \langle \hat{{B}}\rangle \geq  n_d\beta_d + (N -  n_d)\beta_\infty \equiv f(n_d)\;,
\end{eqnarray}
where $\beta_\infty$ is the minimal value of $\langle\hat{\beta} \rangle$ for unbounded local Hilbert space dimension. 
Note that we necessarily have $\beta_d\geq \beta_\infty$, meaning that the above bound $f(n_d)$ is an increasing function of $n_d$.
This directly implies that removing the dimension constraint to more particles result in a decrease in the bound, making it more difficult to violate.
If $\langle \hat{{B}}\rangle<f(n_d)$, then the observed statistics cannot be explained by an ensemble of $N$ parties out of which $n_d$ have dimension at most $d$. 
In other words, a violation of Eq.~\eqref{supp:Bdim} implies that at least $n' = N - n_{d}+1$ parties must have dimension greater than $d$. 
The number of high-dimensional quantum resources certified by the Bell expectation value can be expressed as      
\begin{equation}
    n' = \left\lceil\frac{\langle\hat{B}\rangle-N\beta_d}{\beta_{\infty}-\beta_{d}}\right\rceil \;.
\label{supp:n_d}
\end{equation}
If $\langle \hat{{B}}\rangle = N\beta_\infty$ is attainable (i.e. $N\beta$ saturates the Tsirelson bound of the PIBI), then all parties are required to have a dimension greater than $d$.

For unbounded local Hilbert space dimension, the one-body term gives $\langle\hat{\beta} \rangle\geq \langle(\hat{\pi}_{0|0}+\hat{\pi}_{1|1})(\hat{\pi}_{0|0}+\hat{\pi}_{1|1}-1)+ (\hat{\pi}_{0|1}+\hat{\pi}_{1|0})(\hat{\pi}_{0|1}+\hat{\pi}_{1|0}-1)\rangle\geq (\langle\hat{\pi}_{0|0}\rangle+\langle\hat{\pi}_{1|1}\rangle)(\langle\hat{\pi}_{0|0}\rangle+\langle\hat{\pi}_{1|1}\rangle-1)+ (\langle\hat{\pi}_{0|1}\rangle+\langle\hat{\pi}_{1|0}\rangle)(\langle\hat{\pi}_{0|1}\rangle+\langle\hat{\pi}_{1|0}\rangle-1)\rangle\geq-1/2$, where we used $\langle ^2\rangle\geq\langle \rangle^2$ and that $0\leq\langle\hat{\pi}_{a|\chi} \rangle\leq 1$ for all $a,\chi$.
This implies that $\beta_\infty = -1/2$.

For bounded local Hilbert space dimension $d$, the bound $\beta_d$ can be calculated from a semi-definite program (SDP).
Considering the operator $\hat{\beta}$ as a polynomial of degree two in the non-commuting variables $\hat{\pi} = \{\hat{\pi}_{a|x}\}$, namely $\hat{\beta} = f(\hat{\pi})$, we solve the dimension-constrained polynomial optimization problem 
\begin{equation}
    \beta_d := \begin{array}{ccc} \min_{\hat{\mathbf{\pi}},\ket{\psi}\in \mathbb{C}^d}& \bra{\psi} f(\hat{\pi}) \ket{\psi}  \\
    \mbox{s.t.}& \hat{\pi}_{a|x}\succeq 0, \sum_{a}\hat{\pi}_{a|x} = \mathbb{1}
    \end{array} \;,
    \label{eq:probquditbound}
\end{equation}
which can be relaxed by a sequence of SDPs.
For this, we begin by regarding the measurement settings $\hat{\pi}$ as non-commuting symbols. 
From these operators, we form a list containing all monomials up to some fixed order $l$, namely $\hat{\mathbf{m}} = \{1, \hat{\pi}, \hat{\pi}^2, \ldots, \hat{\pi}^l \}$, where, for instance, $\hat{\pi}^2$ comprises all monomials of degree 2, such as $\hat{\pi}_{0|0}^2$, $\hat{\pi}_{0|0}\hat{\pi}_{11}$, and so on. 
This collection of monomials is then used as a basis for defining the moment matrix $\Gamma = \langle \hat{\mathbf{m}}^\dagger \hat{\mathbf{m}}\rangle$, where every element is given by the expectation value of a corresponding monomial. 
By construction, the moment matrix is Hermitian, $\Gamma = \Gamma^\dagger$, and positive semidefinite, $\Gamma \succeq 0$. 
The objective function $\langle f(\hat{\pi})\rangle$ can therefore be written as a linear combination of the entries of $\Gamma$, i.e., $\langle f(\hat{\pi})\rangle = \mathrm{Tr}(C\Gamma)$, where $C$ is a matrix of real coefficients. 
In the same way, the PSD constraints $\{\hat{\pi}_{a|x}\succeq 0 , \sum_{a}\hat{\pi}_{a|x} = \mathbb{1} \}$ can be recast as linear matrix inequalities $\Lambda(\Gamma)\succeq 0$, where $\Lambda$ denotes the associated hermiticity-preserving linear map.
Thus, the problem we want to solve takes now the SDP form   
\begin{equation}
    \begin{array}{crl} \min_{\Gamma\succeq 0}& \mathrm{Tr}(C\Gamma) \\
    \mbox{s.t.}& \Lambda(\Gamma)\succeq 0
    \end{array} \;.
    \label{eq:SDPpoly}
\end{equation}
It is worth noting that, in the optimization problem Eq.~\eqref{eq:SDPpoly}, the local Hilbert space dimension $d$ has not yet been specified, meaning that it remains unrestricted. As a result, the value obtained in this stage provides a lower bound on $\beta_\infty$, and therefore the inequality $\langle\hat{{B}}\rangle\geq n\beta_\infty$ must be satisfied by any data admitting a quantum realization. From this perspective, the construction can be viewed as a single-particle relaxation of the NPA hierarchy \cite{NavascuesNJP2008}. 
For the Bell inequality considered here, the resulting bound is $\beta = -1/2$, in agreement with the analytical value derived in the previous paragraph. 
The subsequent step is thus to incorporate the constraints associated with the local Hilbert space dimension $d$.
For this purpose, we adopt the Navascués–Vértesi hierarchy introduced in Ref.~ \cite{navascues2015bounding} and implemented via the public package qdimsum.
As a first step, for a fixed local dimension $d$, we generate qudit POVMs together with a state $\hat{\rho}$ and compute the associated moment matrix $\Delta_1$. This procedure is repeated until one obtains a basis of moment matrices $\{\Delta_i\}$ compatible with qudit data. 
The additional condition $\Gamma = \sum_i g_i \Delta_i$, where $\{g_i\}$ are free real coefficients, then confines $\Gamma$ to the intersection between the PSD cone and the subspace corresponding to data subject to the dimensional constraint. 
When this approach is applied to our Bell inequality in the qubit case $d=2$, it reproduces the variational bound $\beta = -1/4$ already at level $l=3$ of the \rev{NPA} hierarchy.
From our results, we conclude that data originating from an $N$-qubit system must satisfy
\begin{equation}
B= \tilde{\mathcal{P}}_{0} + \tilde{\mathcal{P}}_{00} - 2\tilde{\mathcal{P}}_{01}\geq -\dfrac{N}{4} \;.
\label{eq:dimwit2}
\end{equation}
Therefore, if this inequality is violated one must conclude that the system not only has Bell correlations but also that these correlations must involve higher dimensional constituents, i.e. at least qutrits.

\subsection{Bell correlation witness: derivation of Eq.~\eqref{W}}

We derive here the explicit form of the Bell operator $\hat B$ in terms of collective spin--nematic observables for a system of $n$ spin-1 particles.
We consider two local observables labeled by $\chi\in\{0,1\}$.
\begin{equation}
\label{supp:meas}
\hat m_0 = c\,\hat q_{yz} + s\,\hat q_{xy}, \qquad
\hat m_1 = c\,\hat q_{yz} - s\,\hat q_{xy}, \qquad 
\end{equation}
which are linear combinations of quadrupole operators, where $c=\cos\theta$ and $s=\sin\theta$. The local quadrupole operators are defined as 
\begin{equation}
    \hat{q}_{\mu\nu} = \left\lbrace\hat l_\mu,\hat l_\nu \right\rbrace
- \frac{4}{3}\delta_{\mu\nu}\,\mathbb 1.
\end{equation}
where $\hat l_{\mu}$, $\mu = x,y,z$ are the spin-1 operators. 

The corresponding local POVM elements are

\begin{equation}
\label{eq:proj}
\hat\pi_{-1|\chi}=\frac{\hat m_\chi^2+\hat m_\chi}{2} \\ 
\qquad
\hat\pi_{1|\chi}=\frac{\hat m_\chi^2-\hat m_\chi}{2}  .
\end{equation}

We define collective measurement operators
\begin{equation}
\label{supp:MOOmega}
\hat M_\chi=\sum_{i=1}^N \hat m_\chi^{(i)},\qquad \hat{O}_{\chi\gamma} = \frac{1}{2}\sum_{i}\hat{m}_{\chi}^{(i)}\hat{m}_{\gamma}^{(i)}+\mathrm{h.c.} \;, \qquad
\hat\Omega_{\chi\gamma} = \dfrac{1}{2} \sum_{i}(\hat{m}_{\chi}^{(i)})^2(\hat{m}_{\gamma}^{(i)})^2+\mathrm{h.c.}
\end{equation} 
For the projectors defined in Eq.~\eqref{eq:proj} we have
\begin{equation}
\begin{split}
\hat{\Pi}_{-1|\chi} = \dfrac{\hat O_{\chi\chi} + \hat M_{\chi}}{2} \quad&\quad \hat{\Pi}_{1|\chi} = \dfrac{\hat O_{\chi\chi} - \hat M_{\chi}}{2},
\end{split}
\end{equation}
which allows us to express the quadratic terms in the Bell operator Eq.~\eqref{supp:BellOpPi} as: 
\begin{equation}
    (\hat{\Pi}_{-1|0}-\hat{\Pi}_{1|1})^2 + (\hat{\Pi}_{-1|1}-\hat{\Pi}_{1|0})^2  =  ((\hat{M}_0 + \hat{M}_1)^2 + (\hat{O}_{00}-\hat{O}_{11})^2)/2
\end{equation}
We now consider the measurement settings in Eq.~\eqref{supp:meas} and express all terms in the previous equation in the standard spin-1
SU(3) basis consisting of collective spin operators $\hat L_\mu=\sum_i \hat l_\mu^{(i)}$, with $\mu\in\{x,y,z\}$ a cartesian coordinate, and collective nematic (quadrupole) operators
\begin{equation}
\hat Q_{\mu\nu}=\sum_{i=1}^N {\hat{q}}_{\mu\nu}^{(i)}.
\end{equation}

From the definition of $\hat m_\chi$ we obtain
\begin{equation}
\hat M_0+\hat M_1 = 2c\,\hat Q_{yz},
\qquad
\frac{(\hat M_0+\hat M_1)^2}{2}=2c^2\hat Q_{yz}^2.
\end{equation}
Similarly, using
$\hat m_0^2-\hat m_1^2=2cs\{\hat q_{yz},\hat q_{xy}\} = -2cs \hat{q}_{xz}$ we find 
\begin{equation}
\hat O_{00}-\hat O_{11}=-2cs\,\hat Q_{xz},
\qquad
\frac{(\hat O_{00}-\hat O_{11})^2}{2}=2s^2c^2\,\hat Q_{xz}^2.
\end{equation}
Let us now consider the remaining single-particle term in the Bell operator Eq.~\eqref{supp:BellOpPi}. We first express it in terms of the collective operators Eq~\eqref{supp:MOOmega}:
\begin{equation}
    2(\hat\Pi_{-11|01}+\hat\Pi_{-11|10}) = \hat{\Omega}_{01} -\hat{O}_{01}.
\end{equation}
As before, we express it in terms of the spin and quadrupole basis using measurement settings in Eq.~\eqref{supp:meas}.
\begin{equation}
\hat{\Omega}_{01} -\hat{O}_{01} = s^2c^2(\hat{Q}_{zz}+(\hat{Q}_{xx}-\hat{Q}_{yy})/2-2N/3) + s^2(1+s^2)(\hat Q_{zz}/2 + 2N/3)
\end{equation}

Collecting all contributions yields the exact operator identity
\begin{equation}
\hat B(\theta)
=2c^2\hat Q_{yz}^2
+s^2c^2\!\left(2\hat Q_{xz}^2+\hat{Q}_{zz}+(\hat{Q}_{xx}-\hat{Q}_{yy})/2-2N/3\right)
+s^2(1+s^2)(\hat Q_{zz}/2 + 2N/3).
\end{equation}
Finally, choosing $\theta=\pi/4$, one arrives at the witness
\begin{equation}
{B} = \left\langle\hat B\left(\frac{\pi}{4}\right)\right\rangle
=\langle \hat Q_{yz}^2\rangle
+\frac{1}{2}\langle \hat Q_{xz}^2\rangle
+\frac{1}{8}\langle \hat Q_{xx}-\hat Q_{yy}+5\hat Q_{zz}\rangle
+\frac{N}{3} \geq -\frac{N}{4} .
\end{equation}

Note that a similar witness can be derived by exchanging $y\leftrightarrow x$ in the above calculation. This gives
\begin{equation}
{B}' =
\langle \hat Q_{xz}^2\rangle
+\frac{1}{2}\langle \hat Q_{yz}^2\rangle
+\frac{1}{8}\langle \hat Q_{yy}-\hat Q_{xx}+5\hat Q_{zz}\rangle
+\frac{N}{3} \geq -\frac{N}{4} \;.
\end{equation}
Since in our experiment the axis $x, y$ are randomly defined at each shot (see main text), it is convenient to consider a witness symmetric under this relabeling.
Due to convexity, we can write
\begin{equation}
   \frac{ {B} + {B}'}{2} =\frac{3}{4}( \langle \hat Q_{yz}^2\rangle+\langle \hat Q_{xz}^2\rangle) + \frac{5}{8}\langle \hat{Q}_{zz} \rangle + \frac{N}{3} \geq -\frac{N}{4}
\end{equation}

Defining $\overline{B}\equiv\frac{B+B'}{2}$, the collective witness is related to the Bell expression by
\begin{equation}
    W=\langle \hat W \rangle = 2+\frac{4}{N}\overline{B} = \frac{3}{N}( \langle \hat Q_{yz}^2\rangle+\langle \hat Q_{xz}^2\rangle) + \frac{5}{2N}\langle \hat{Q}_{zz} \rangle + \frac{10}{3}, 
\end{equation}
which is Eq.~\eqref{W} in the main text.Therefore, the Bell-local bound $\overline{B}\ge0$ implies $W\ge2$, whereas the qubit bound $\overline{B}\ge-N/4$ implies $W\ge1$. Thus, $W<2$ certifies Bell correlations, while $W<1$ additionally excludes every $N$-qubit realization.

\section{Experimental methods}
\subsection{Initial state preparation}
We prepare a Bose-Einstein condensate (BEC) of approximately $31000$ $^{87}$Rb atoms in the $|F=1, m_F=0\rangle$ state, confined within a crossed optical dipole trap with trapping frequencies of $(\omega_x, \omega_y, \omega_z) = 2\pi \times (163, 77, 100)\,\mathrm{Hz}$. The bias magnetic field is actively stabilized at $0.99\,\mathrm{G}$ via a feedback control loop, yielding a quadratic Zeeman shift of $q_{\rm B} = 2\pi \times 70.6\,\mathrm{Hz}$ or $19.9|c_2|$. This bias field magnitude is chosen to suppress environmental radio-frequency (rf) noise, which deleteriously couples the three Zeeman states at low magnetic field and leads to decoherence~\cite{liu2022nonlinear}. Initially, the atomic levels are dressed by a microwave field red-detuned by $10\,\mathrm{MHz}$ from the $|1,0\rangle \leftrightarrow |2,0\rangle$ transition. This induces a large effective quadratic Zeeman shift $q_{\mathrm{MW}}$, ensuring that the total shift $q = q_{\rm B} + q_{\mathrm{MW}}$ is significantly larger than $|c_2|$, thereby maintaining the atoms in the polar state. To initiate spin-nematic squeezing, the microwave frequency is quenched to a blue detuning of $7\,\mathrm{MHz}$. This flips the sign of $q_{\mathrm{MW}}$ to partially compensate for $q_{\rm B}$. By calibrating the microwave power, the net shift $q$ is tuned to match $|c_2|$. Prior to the squeezing dynamics, we improve the purity of the initial state by applying additional microwave pulses to transfer residual atoms in the $|1, \pm 1\rangle$ states to the $|2, \pm 2\rangle$ states. These impurities are subsequently removed from the trap using a resonant probe beam. The final atom number is $31000$ with a standard deviation of $380$, and the spin-dependent interaction strength is calibrated to be $c_2 = -2\pi \times 3.55\,\mathrm{Hz}$, following the procedure established in Ref.~\cite{Guo:2021aa}.\\
\noindent
\subsection{Measurement of $\hat Q(\theta)$}
We implement the following protocol to measure the spin-nematic quadrature operator $\hat{Q}(\theta)$: (1) We first perform a rotation within the spin-nematic plane via encoding a spinor phase $\theta=\theta_{1}+\theta_{-1}-2\theta_0$, where $\theta_i$ indicates the phase encoded in $|m_F=i\rangle$ mode, to map the target observable $\hat{Q}(\theta)$ onto $\hat{L}_x$. (2) To extract the information of $\hat{L}_x$, we apply a $\pi/2$ pulse around $\hat{L}_y$ axis to map $\hat{L}_x$  onto the $\hat{L}_z$. Thus, the final readout of $\hat{L}_z$ yields the value of $\hat{Q}(\theta)$ of the original state. Note that since the spin-nematic squeezed state preserves rotational symmetry about the $\hat{L}_z$ axis, measuring $\hat{Q}(\theta)$ is statistically equivalent to measuring any quadrature component $\cos(\psi)\hat{Q}(\theta) + \sin(\psi)\hat{Q}'(\theta)$ within the equatorial plane $Q_{xz}$-$Q_{yz}$.
\noindent
\subsection{Spin rotation in the spin-nematic plane}
To implement the rotation within the spin-nematic plane, we utilize the following transformations:
\begin{equation}
\begin{aligned}
\mathrm{e}^{-\mathrm{i}\hat{N}_0\theta}\hat{L}_x\mathrm{~e}^{\mathrm{i}\hat{N}_0\theta} = \hat{L}_x\cos\theta - \hat{Q}_{yz}\sin\theta, \\
 \mathrm{e}^{-\mathrm{i}\hat{N}_0\theta}\hat{Q}_{yz}\mathrm{~e}^{\mathrm{i}\hat{N}_0\theta} = \hat{Q}_{yz}\cos\theta + \hat{L}_x\sin\theta.
 \end{aligned}
 \end{equation}
These relations show that adjusting the relative phase between the $|1,0\rangle$ state and the $|1, \pm 1\rangle$ states is equivalent to performing a rotation in the spin-nematic plane. Experimentally, we apply a microwave field with a power of $9\,\mathrm{W}$, blue-detuned by $4\,\mathrm{MHz}$ from the $|1,0\rangle \leftrightarrow |2,0\rangle$ transition, which boosts the quadratic Zeeman shift $q$ to $2\pi\times 123\,\mathrm{Hz}$ ($34.6 |c_2|$). Such large energy shift strongly suppresses the spin-mixing interactions, rendering the quadratic Zeeman term $-q\hat{N}_0$ the dominant contribution to the Hamiltonian. Under this condition, the time-evolution propagator over a duration $\tau$ is given by $\hat{U}(\tau)=\exp(\mathrm{i}\hat{N}_0 q\tau)$. By precisely controlling the evolution time $\tau$, we can realize the desired rotation with an angle of $\theta = -q\tau$.\\
\noindent
\subsection{Calibration of the quadratic Zeeman shift $q$}
A precise measurement of $q$ is required to achieve accurate rotations. The specific calibration procedure is as follows: starting from a polar state, we apply a resonant rf Rabi pulse to transfer approximately $25\%$ of the $|1,0\rangle$ population to the $|1, \pm 1\rangle$ states. The microwave field is then activated to quench the quadratic Zeeman shift to the target value $q_t$. In the regime where $q_t \gg |c_2|$, spin-mixing dynamics is energetically suppressed; consequently, the system undergoes a pure phase interrogation with negligible population variation. After a variable hold time $\tau$, the microwave field is quenched back to initiate spin mixing at $q=|c_2|$ over a short duration of $t=5\,\mathrm{ms}$. Based on the mean field dynamics, the resulting temporal evolution of $\rho_0(t)$ can be approximated as:
$$\rho_0(\tau) - \rho_0 \approx 2c_2 t \rho_0 (1-\rho_0) \sin(2q_t \tau)$$
where $\rho_0 = \rho_0(0) \approx 0.75$ is the initial population of the $|1,0\rangle$ state. By fitting the measured $\rho_0(\tau)$ to this sinusoidal function, we extract the quadratic Zeeman shift $q_t$ to be $2\pi\times123~\mathrm{Hz}$.\\
\noindent
\subsection{Measurement of $\hat W$ for the qutrit case}
Evaluation of the nonlinear Bell witness term $\hat{W}_{\rm non}$ requires knowledge of both $\langle \hat{Q}_{xz}^2\rangle$ and $\langle \hat{Q}_{yz}^2\rangle$. One approach to retrieve these quantities is through the joint measurement, as recently demonstrated in Ref.~\cite{cao_joint_2025}. The key idea is to map the two squeezed modes $\hat{a}_{\pm}=(\hat{a}_1\pm\hat{a}_{-1})/\sqrt{2}$ onto the Zeeman modes $\hat{a}_{\pm1}$, which can be simultaneously measured via interference with two independent local oscillators. While such joint measurement ideally preserves the spin-nematic squeezing and yields a faithful estimation of the nematic tensor operators, this protocol is in practice susceptible to pulse errors and magnetic field noise, both of which severely degrade the attainable squeezing parameter and pose significant challenges for observing a Bell violation. 

Fortunately, joint measurement proves unnecessary for evaluating the Bell witness. Considering that $\hat{Q}_{yz}$ and $\hat{Q}_{xz}$ are related by a spin rotation about the $\hat{L}_z$ axis, we define the generalized quadrature operator $\hat{\mathcal{Q}}\equiv\hat{Q}_{yz}\cos\phi+\hat{Q}_{xz}\sin\phi$, where $\phi$ denotes the relative phase between the atomic Larmor phase and the rf phase. For spin-mixing dynamics starting from the polar state, the atomic Larmor phase is randomized and uniformly distributed over $[0, 2\pi]$ owing to the rotational symmetry about the $\hat{L}_z$ axis, and is uncorrelated with the rf phase. Consequently, in each experimental cycle, $\hat{\mathcal{Q}}$ is mapped onto $\hat{L}_z$ with a random $\phi$ prior to measurement. The cycle-averaged expectation of $\hat{L}_z^2$ is then given by $\overline{\langle\hat{\mathcal{Q}}^2\rangle}=(1/2\pi)\int_0^{2\pi}\langle(\hat{Q}_{yz}\cos\phi+\hat{Q}_{xz}\sin\phi)^2\rangle d\phi=\langle\hat{Q}_{yz}^2+\hat{Q}_{xz}^2\rangle/2$, from which the nonlinear witness term is expressed as $\langle\hat{W}_{\rm non}\rangle=(6/N)\langle\overline{\hat{\mathcal{Q}^2}\rangle}$, where $\overline{\cdot}$ denotes the statistical average over $\phi$. This approach requires no complicated pulse sequence as in the joint measurement, and is considerably more robust against experimental imperfections.

Another subtlety warranting extra attention concerns the mean value of $\hat{\mathcal{Q}}$. For a unitary spin-nematic squeezed state, we expect $\langle\hat{\mathcal{Q}}\rangle=0$, thus $\langle\hat{\mathcal{Q}}^2\rangle=\Delta^2\hat{\mathcal{Q}}$ and $\hat{W}_{\rm non}$ is directly related to the squeezing parameter. In our experiment, however, we observe a non-zero offset that is independent of $\phi$ and remains nearly constant over timescales of a few days, yet exhibits a long-term fluctuation within the range of $(-100,100)$. We attribute this offset to the imperfect atom number detection, whose precision is limited by inhomogeneities of the probe light intensity and shot-to-shot variations in the position of the atomic ensembles. Left uncorrected, this offset effectively inflates $\hat{W}_{\rm non}$, constituting an additional obstacle to the direct observation of Bell correlation. 

To address this issue, we can define $\mathcal{Q}_{r}^{(i)}\equiv\mathcal{Q}^{'(i)}-\mathcal{Q}^{'(i+1)}=\mathcal{Q}^{(i)}-\mathcal{Q}^{(i+1)}$ as the difference between $\mathcal{Q}$ values measured in two consecutive shots. Here, $(i)$ and $(i+1)$ denote the shot indices, and $\mathcal{Q}^{'}$ is the measured value which differs from the true value $\mathcal{Q}$ by a constant offset. The mean squared value of $\mathcal{Q}_r$ can be written as $\overline{\mathcal{Q}_{r}^2}=\sum_i\mathcal{Q}_{r}^{(i)2}/n=\sum_i{(\mathcal{Q}^{(i)}-\mathcal{Q}^{(i+1)})}^2/n=2\sum_i\mathcal{Q}^{(i)2}/n-2\sum_i\mathcal{Q}^{(i)}\mathcal{Q}^{(i+1)}/n$ where $n$ is the total number of shots. Since the quadrature phase $\phi$ is uncorrelated across multiple shots, the cross-term vanishes in the limit of $n\rightarrow\infty$, i.e., $\sum_i\mathcal{Q}^{(i)}\mathcal{Q}^{(i+1)}\rightarrow0$ , yielding $\overline{\mathcal{Q}_{r}^2}\rightarrow2\sum_i\mathcal{Q}^{(i)2}/n=NW_{\rm non}/3$. Physically, this approach is analogue to the four-point measurement scheme employed in cavity QED experiment~\cite{braverman_near-unitary_2019}, which alternately measures $\langle\hat{Q}_{yz}\rangle$ and $-\langle\hat{Q}_{yz}\rangle$ and take their difference to actively cancel the measurement offset. To improve the convergence of the cross-term, we randomly shuffle the original dataset $\{{\mathcal{Q}^{'(1)}},{\mathcal{Q}^{'(2)}},\cdots,{\mathcal{Q}^{'(n)}}\}$ over $100$ times, and take the resulting average of $\overline{\mathcal{Q}_{r}^2}$ as the final $W_{\rm non}$ which is reported in the main text. It is noted that this shuffling may introduce an overestimation of $W_{\rm non}$ in the presence of long-term drift in the offset, which could be mitigated in the future experiments by directly implementing the four-point measurement scheme.\\
\noindent
\subsection{Measurement of $\hat W$ for the qubit case}
The qubit system discussed in this work is constructed from the spin-1 qutrit manifold by selectively depleting the population of a specific superposition state. This effective two-level system preserves the three-dimensional operator structure defined in the Bell witness while adhering to the physical bounds of a qubit. The subspaces spanned by $\{\hat{a}_+,\hat{a}_0\}$ modes and $\{\hat{a}_-,\hat{a}_0\}$ modes form two independent SU(2) sectors, corresponding to distinct spin-nematic Bloch spheres.  Following the generation of spin squeezing via spin-mixing dynamics, our objective is to isolate the qubit subspace spanned by $\{\hat{a}_-,\hat{a}_0\}$ by eliminating the population in the symmetric $\hat a_+$ mode. Consequently, the system becomes polarized within the $\{\hat{a}_+,\hat{a}_0\}$ Bloch sphere, whereas the squeezing information is preserved within the $\{\hat{a}_-,\hat{a}_0\}$ Bloch sphere.

The experimental sequence begins with the generation of spin-nematic squeezing via spin-mixing dynamics. To minimize the effect of magnetic field noise, the optimal squeezing is first aligned along the $\hat{Q}_{xz}$ and $\hat{Q}_{yz}$ axes through a spinor phase encoding. An rf Rabi $\pi/2$ pulse is then applied around the $\hat{L}_x$ axis, which induces the transformations $\hat{a}_0\leftrightarrow\hat{a}_+$ and $\hat a_-\to \hat a_-$. Crucially, this step maps the information of the initial symmetric state $|+\rangle\equiv ({|1,1\rangle+|1,-1\rangle})/\sqrt{2}$ onto the Zeeman sublevel $|1,0\rangle$. Immediately following this rotation, a microwave $\pi$-pulse transfers the population in $|1,0\rangle$ to the $|2,0\rangle$ state, and these atoms are removed from the trap using a resonant probe beam. Finally, a second rf $\pi/2$ pulse around $\hat{L}_x$ is applied to rotate the current $|+\rangle$ component (carrying the initial $|0\rangle$ population) back to $|0\rangle$. 

With the system prepared in the target qubit subspace, the squeezing observables are measured by rotating the fluctuations onto the $\hat{L}_z$ axis. Specifically, the measurement of $\hat{Q}_{xz}$ is achieved by applying an rf $\pi/2$ pulse around the $\hat{L}_x$ axis after being converted to $\hat{L}_y$ via a spinor phase encoding , while the measurement of $\hat{Q}_{yz}$ is performed using an rf $\pi/2$ pulse around the $\hat{L}_y$ axis after being converted to $\hat{L}_x$. Other linear terms in the Bell witness are determined using the standard procedures described in the previous sections.\\
\noindent
\subsection{Accuracy of the atom number counting}
\begin{figure}[ht]
    \centering    \includegraphics[width=0.7\columnwidth]{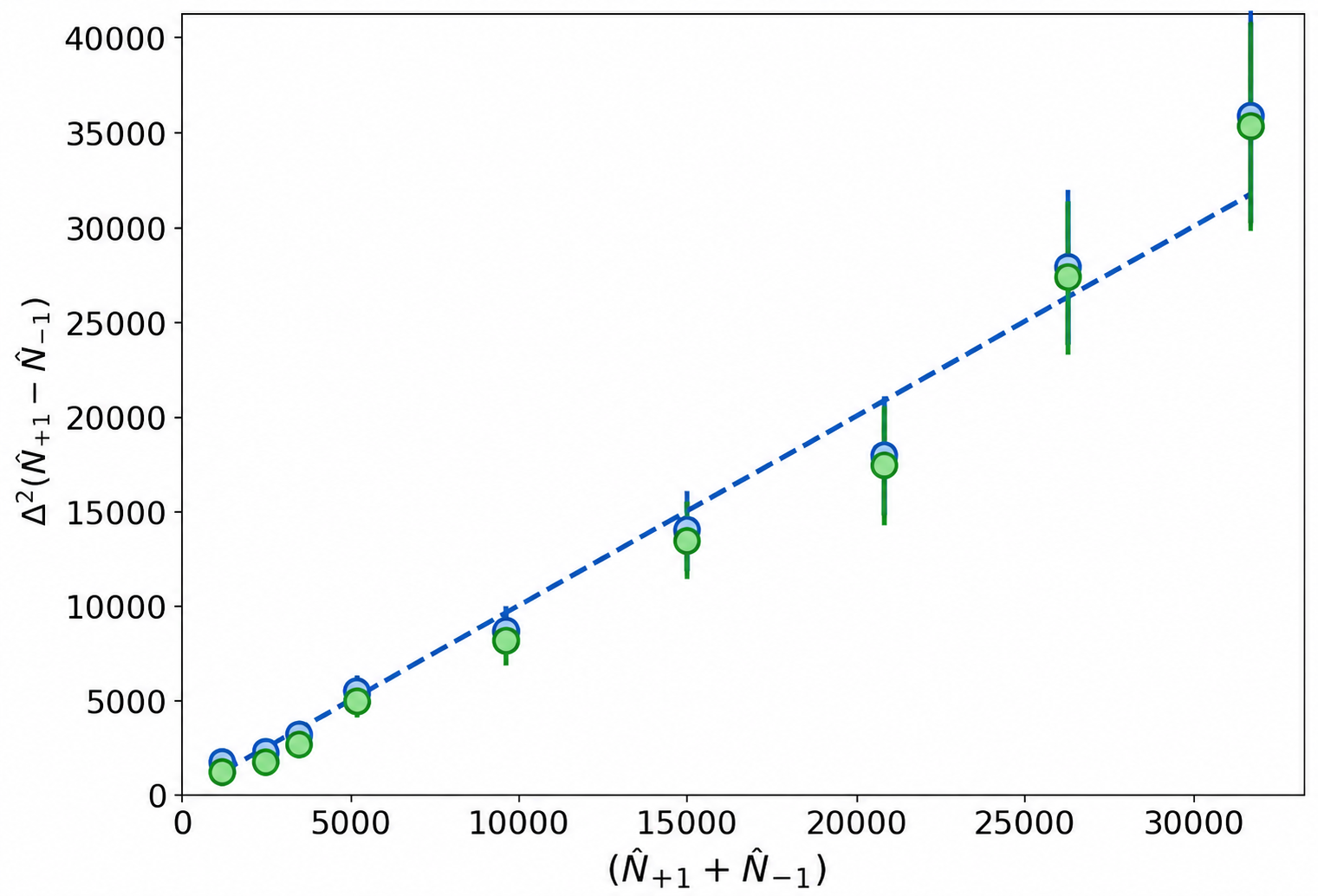}
    \caption{\textbf{ Calibration of the atom number counting accuracy}. The blue (green) circles denote the measured variance of the number difference between $|1, \pm 1\rangle$ modes before (after) subtracting atom-independent detection noise of $(\Delta \hat{L}_z)_{\rm det} = 23.1$, obtained from $100$ independent experiments respectively.}
    \label{atom_num_counting}
\end{figure}

The atom number counting is calibrated with the methods introduced in \cite{Luo17deterministic}. To verify the counting accuracy, we apply a resonant $\pi/2$ rf pulse to the polar state $|0\rangle^{\otimes N}$, which prepares atoms in the coherent superposition state $\left(\frac{|1\rangle+|-1\rangle}{\sqrt{2}}\right)^{\otimes N}$. With correct atom number counting, the variance of $(\hat{N}_1-\hat{N}_{-1})$ should correspond to the quantum projection noise given by $\langle\hat{N}_1+\hat{N}_{-1}\rangle$, which is consistent with the measurement results shown in Fig.~\ref{atom_num_counting} , thereby confirming the accuracy of the detection. 

\clearpage

\end{document}